%% file: ms7.tex
\documentclass[12pt,preprint]{aastex}

\newcommand{\bdv}[1]{\mbox{\boldmath$#1$}}

\def\au{{\rm au}} 
 
\def\kms{{\rm km}\,{\rm s}^{-1}}
\def\masyr{{\rm mas}\,{\rm yr}^{-1}}
\def\kpc{{\rm kpc}}
\def\mas{{\rm mas}}

\def\max{{\rm max}}
\def\min{{\rm min}}
\def\rel{{\rm rel}}

\def\rot{{\rm rot}}

\def\e{{\rm E}}
\def\bpi{{\bdv\pi}}
\def\bmu{{\bdv\mu}}

\def\bomega{{\bdv\omega}}

\def\btheta{{\bdv\theta}}

\def\bv{{\bf v}}

\begin{document}
\title{{\it Roman} FFP Revolution: Two, Three, Many Plutos}

\input author.tex

\begin{abstract}

{\it Roman} microlensing stands at a crossroads between its originally
charted path of cataloging a population of cool planets that has subsequently
become well-measured down to the super-Earth regime,
and the path of free-floating planets
(FFPs), which did not even exist when {\it Roman} was chosen in 2010,
but by now promises revolutionary insights into planet formation and
evolution via their possible connection to a spectrum of objects
spanning 18 orders of magnitude in mass.  Until this work, it
was not even realized that the two paths are in conflict: {\it Roman}
strategy was optimized for bound-planet detections, and FFPs
were considered only in the context of what could be learned about
them given this strategy.  We derive a simple equation that
mathematically expresses this conflict and explains why the current
approach severely depresses detection of 2 of the 5 decades of potential FFP
masses, i.e., exactly the two decades, $M_{\rm Pluto}< M <
2\,M_{\rm Mars}$, that would tie terrestrial planets to the
proto-planetary material out of which they formed.  FFPs can be either
genuinely free floating, or they can be bound in ``Wide'', ``Kuiper'',
and ``Oort'' orbits, whose separate identification will allow further
insight into planet formation.  In the (low-mass) limit that the source
radius is much bigger than the Einstein radius, $\theta_* \gg \theta_\e$, the
number of significantly magnified points on the FFP light curve is
$$
N_{\rm exp} = 2{\Gamma\theta_*\sqrt{1-z^2}  \over \mu_\rel}=
3.0\biggl({\Gamma \over 4/\rm hr}\biggr)
\biggl({\theta_* \over 0.3\,\mu\rm as}\biggr)
\biggl({\mu_\rel \over 6\,\mas/\rm yr}\biggr)^{-1}
\biggl({\sqrt{1-z^2}\over 0.86}\biggr),
$$
where the lens-source proper motion $\mu_\rel$, the source impact
parameter $z$, and $\theta_*$ are scaled to their typical values,
and the cadence $\Gamma$ is normalized to the value chosen to optimize
the original {\it Roman} microlensing goals.
Hence, the typical number of significantly magnified points on an FFP
light curves is $N_{\rm exp} = 3.0$, whereas $N=6$ are needed for an
FFP detection. Thus, unless $\Gamma$ is doubled, FFP detection will be
driven into the (large-$\theta_*$, small-$\mu_\rel$) corner of
parameter space, reducing the detections by a net factor of 2 and
cutting off the lowest-mass FFPs.

\end{abstract}

\keywords{gravitational lensing: micro}

\section{{Introduction}
  \label{sec:intro}}

\subsection{The Waning Potential of Bound Planets}
\label{sec:waning}

At the time that a ``Wide Field Imager in Space for Dark Energy and
Planets'' was proposed \citep{gould09} to the 2010 Decadal Committee
and was later adopted by the National Research
Council\footnote{Astro2010: The Astronomy and Astrophysics Decadal
Survey; New Worlds, New Horizons in Astronomy and Astrophysics;
https://science.nasa.gov/astrophysics/resources/decadal-survey/astro2010-astronomy-and-astrophysics-decadal-survey}
as the Wide-Field Infrared Space Telescope ({\it WFIRST}),
microlensing planets were being discovered at the rate of a few per
year.  In that context, the resulting homogeneous sample of
$\cal{O}$(1000) microlensing planets, over the full range of masses,
in the otherwise unreachable cold, outer regions of solar systems,
would indeed be a ``revolution'' by completing the systematic census
of exo-planets, which had been pioneered in the warm and hot regions
by radial-velocity (RV) and transit studies, respectively.  Moreover,
in contrast to the ground-based detections, which delivered only
planet-host mass-ratio ($q$) measurements, a substantial fraction of
{\it WFIRST} planets would yield host-mass ($M_{\rm host}$) measurements,
and thereby also planet-mass ($M_{\rm planet}= q M_{\rm host}$)
measurements.

Fast forward 15 years, and these formerly ``revolutionary'' prospects
have begun substantially merging into the mainstream.  Based on a
systematic analysis \citep{kmt4yr} of the first four years (2016-2019)
of KMTNet \citep{kmtnet} microlensing, there are already about 200 KMTNet
planets in a homogeneous sample (2016-2024), and this will likely increase
to about 300 by the time that {\it WFIRST} (now renamed {\it Roman})
is launched.  As soon as adaptive optics (AO) are available on
extremely large telescopes (ELTs), it will be possible to make mass
measurements of the majority of hosts (and therefore planets) in the
2016-2019 KMTNet sample, and the planets detected in later years will also
gradually become amenable to measurement as the source-lens separations
continue to increase \citep{masada}.

Certainly, {\it Roman} will detect several times more planets than
KMT.  Moreover, late-time AO observations will enable mass
measurements for an even larger fraction of {\it Roman} planets than
KMT planets, simply because its
sources are systematically fainter\footnote{For example, according to
Figure~9 of \citet{masada}, about 20\% of KMT planetary-microlensing sources are
giants, which probably require source-lens separations of 10 FWHM
($140\,\mas$ for a 39m telescope) to allow for the 10 magnitude
(factor 10000) contrast ratios that are needed to probe down most of the
main sequence.  The wait time for typical lens-source relative proper motions
$\mu_\rel\sim 6\,\masyr$ would be about 25 years.
By contrast, only a tiny fraction of {\it Roman} planetary-microlensing sources
will be giants.}.
However, in most respects, this
will be an evolutionary, not revolutionary, development.
The one remaining revolutionary element of the original {\it
  WFIRST/Roman} plan is its potential to probe the planet mass-ratio
function well below what has been achieved from the ground.

\subsection{{A Spectrum Haunts Microlensing: the FFP Mass Spectrum}
  \label{sec:spectrum}}

On the other hand, the fading revolutionary potential of the original
{\it WFIRST/Roman} program has been more than matched by the surging
potential of its application to free-floating planets (FFPs) and
wide-orbit planets \citep{yeegould23}.

Observationally, FFPs are short single-lens single-source (1L1S)
microlensing events.  These could indeed be unbound to any host, but
may also be due to wide-orbit planets whose hosts are too far away to
leave any trace on the event.  As discussed by \citet{gould16}, FFP
events can be resolved into moderately wide (hereafter: Wide), Kuiper,
Oort, and Unbound objects using late-time high-resolution
observations.  As we will show, for the great majority of space-based
FFP discoveries, these four categories can be distinguished within 10
years after the microlensing event, but at the time of discovery,
there will be at most hints as to whether they are actually unbound
(``free'') or they are bound.  And most often, there will not even be
hints.  Hence, in the present work, we keep the nomenclature ``FFP''
as an {\it observational} classification of objects whose physical
nature must still be determined on an event-by-event basis.

\citet{sumi11} were the first to propose a large FFP population, a year too
late to be considered by the Astro2010 report as part of the {\it WFIRST}
mission.  While \citet{mroz17} did not confirm 
the \citet{sumi11} Jupiter-mass FFP population, they did find evidence for
a large FFP population of much lower mass based on the detection of six
1L1S events with short Einstein timescales, $t_\e<(1/3)\,$d.  These were all
point-source point-lens (PSPL) events.  Subsequently,
several studies led to the detection of nine additional FFPs, all
finite-source point-lens (FSPL) events, which therefore yielded measurements
of their Einstein radius, $\theta_\e\propto\sqrt{M}$
\citep{ob161540,ob121323,ob190551,ob161928,kb172820,kb192073,moaffp,kb232669},
all with $\theta_\e\la 9\,\mu$as.

While both \citet{sumi11} and \citet{mroz17} expressed the ensemble of
their detections of short-$t_\e$ PSPL events in terms of simplified
$\delta$-function FFP mass functions, \citet{gould22} used the four
$\theta_\e<9\mu$as FSPL detections from their own study,
combined with their study's absence
of FSPL events in the ``Einstein Desert'' ($9\,\mu{\rm as}\la
\theta_\e \la 25\,\mu{\rm as}$), to derive a power-law FFP mass
function.  They showed that this mass function was consistent with the
six PSPL events reported by \citet{mroz17}.  They also showed that if
this, albeit crudely measured, power law were extended by 18 orders of
magnitude, it was consistent with the previous detections of
interstellar asteroids and comets. 

Thus, \citet{gould22} both confirmed the earlier suggestions of
\citet{sumi11} and \citet{mroz17} that there were substantially more
FFPs than stars, but also tied these objects to the potentially vast
population of very small planets, dwarf planets, and sub-planetary
objects, either analogs of Kuiper-Belt and Oort-Cloud objects or
potentially ejected from their solar systems.

Indeed, there is already important, if still suggestive, evidence for
a very large population of sub-Earth-mass objects.
Compared to most of the rest of the current sample
of FFPs, OGLE-2016-BLG-1928 \citep{ob161928} has an unusually small
$\theta_\e=0.84\,\mu$as.  Considering that,
\begin{equation}
  \theta_\e\equiv \sqrt{\kappa M\pi_\rel}=
  0.84\biggl({M \over M_\oplus}\biggr)^{1/2}
  \biggl({\pi_\rel \over 29\,\mu\rm as}\biggr)^{1/2}; 
  \qquad \kappa\equiv {4G\over c^2\au}\simeq 8144\,{\mu{\rm as}\over M_\odot},
  \label{eqn:thetaedef}
\end{equation}
this could in principle be a few-Earth-mass FFP in the Galactic bulge,
for which the lens-source relative parallax is usually in the range
$\pi_\rel\la 30\,\mu$as.  However, by chance, this scenario can be
virtually ruled out by the event's high observed relative proper
motion $\mu_\rel = 10.6\,\masyr$, and the measured vector proper
motion of the source, which together are strongly inconsistent with
bulge kinematics for the lens.  See their Figure~3.  For a typical
disk $\pi_\rel\sim 60\,\mu$as, this object would have mass $M\sim
0.5\,M_\oplus$.

Moreover, one of the two FFPs found by \citet{moaffp} has
$\theta_\e=0.90\,\mu$as.  While, in contrast to OGLE-2016-BLG-1928, there are no
constraints for this object on whether it resides in the disk or the bulge,
the high-frequency (i.e., two!) of FFPs with $\theta_\e\la 1\,\mu$as,
despite the severe difficulty of detecting them, suggests an intrinsically
high frequency of low-mass FFPs.

The physical origins of FFPs, whether each is ultimately identified as
being bound or unbound, and whether it is of low or high
mass, are likely tied together by a single history of planet formation
and early dynamical evolution.  Thus, detailed statistical studies of these
objects, ultimately broken down into relatively Wide, Kuiper, Oort, and
Unbound, will shed immense light on the process of planet
formation and evolution.  In particular, for the bound subsample, it will
be possible to measure the masses and physical host-planet projected
separations on an object-by-object basis, as was already discussed by
\citet{gould16}, and which we will further discuss below.  Hence,
it will also be possible to study the differences in the mass functions
of these different sub-populations, which will be critical input for
theories of planet formation.  Individual
masses for genuinely free-floating objects will be more challenging,
and we will also discuss these challenges.

However, the main point from the perspective of this introduction is
that in the 15 years since the 2010 Decadal process, the issue of the
FFP mass function (or mass functions at different levels of host separation)
has emerged from absolutely nothing, to weak sister of the more recognized
question of a bound-planet census, to intriguing question of the day,
to a unique probe of planet formation and evolution that links planets
  with protoplanetary objects.
Yet, as in most proto-revolutionary situations, understanding
of the spectacular emergence of this new field and new questions has
lagged dangerously behind actual developments.

In particular, at the moment, the final observation strategy is being
formulated for {\it Roman} based primarily on bound-planet yield, and
the only role of FFPs in this process is to verify that a relatively
weak Level 1 requirement on FFPs can be met given whatever strategy is
adopted in pursuit of bound planets.

By  prioritizing revolutionary FFP science, this paper truly turns
the entire process on its head.

\subsection{{Two Approaches to the FFP Revolution}
  \label{sec:two-tactics}}

One approach is to consider what {\it Roman} can achieve for FFPs
according to its adopted strategy.  There have been two major studies
of the detection and characterization of FFPs by {\it Roman}, which
individually and collectively provide valuable insights on the
measurement process.  \citet{johnson20} studied {\it Roman} detections
of FFPs at a range of masses.  For example, they considered an FFP
mass function that is flat at $2\,{\rm dex}^{-1}$ (per star) for $M<5.2 M_\oplus$
and falls with a $p = -0.73$ power law above that value.  This was
necessarily arbitrary because there were no published estimates of the
FFP mass spectrum at that time.  From the present standpoint, the fact
that the mass function is flat at low masses will simplify the
interpretation of their results.

Their Figure~5 and Table~2 show that there are 5.0 times more
detections at $M=1\,M_\oplus$ (``Earth'') than at $M=0.1\,M_\oplus$
(``Mars'') despite the fact that their adopted mass function assigns
equal frequency to each class.  At first sight, this seems natural because
smaller masses generate shorter and/or weaker perturbations, which are
harder to detect.  And their Figure~11 seems to confirm this by
showing that the source stars for detected events are systematically
brighter by about 2 magnitudes for the Mars-class than Earth-class objects,
seemingly because (according to one's first instinct) brighter
source flux is required for the former.  Table~2 similarly
shows that $M=0.01\,M_\oplus$ (``Moon'') class objects are nearly
impossible to detect unless they are extraordinarily numerous.

\citet{johnson22} investigate the seemingly intractable degeneracies
in the basic microlensing parameters $(t_0,u_0,t_\e,\rho,f_s)$ in the
large-$\rho$ limit, which generally applies to the lowest-mass FFPs.
Here $(t_0,u_0,t_\e)$ are the basic \citet{pac86} parameters:
the time of maximum, the impact parameter (scaled to $\theta_\e$), and
the Einstein timescale $t_\e=\theta_\e/\mu_\rel$, while
$\rho\equiv\theta_*/\theta_\e$ is the ratio of the angular source radius
to the Einstein radius, and $f_s$ is the source flux.
For example, in this limit, there is an almost perfect degeneracy between
$f_s$ and $\rho$ because the observed excess flux is $\Delta F = 2 f_s/\rho^2$.
There are several other degeneracies as well.

Here we build on these studies 
by taking the opposite approach: we ask what can be done to
resolve the problems they identify, notably the poor sensitivity to
low-mass FFPs, by altering the {\it Roman} strategy.

In this quest, we begin by ignoring any
constraints arising from the ``official goal'' of {\it Roman} to
``complete the census of planets''.  Subsequently, recognizing that even
the most successful revolutions must eventually come to terms with the ``old
order'', we ask what compromises can be made to reconcile these two somewhat
conflicting goals.

For a primer on microlensing as it specifically relates to FFP microlensing
events and mass measurements, see Appendix~\ref{sec:primer}.

In Section~\ref{sec:killing}, we show analytically that two
decades of low-mass FFP detections are being ``killed off'' by
{\it Roman}'s low adopted survey cadence of $\Gamma = 4\,{\rm hr}^{-1}$.
In Section~\ref{sec:thetae}, we show how the critical Einstein-radius
parameter, $\theta_\e$, can be measured for large-$\rho$ (low-mass) FFPs,
despite the
fact that very few will have source color measurements, which is usually
considered a sine qua non for such measurements.
In Section~\ref{sec:bound}, we discuss two issues that are specifically related
to bound FFPs, including that all FFP detections must be subjected to late-time,
high-resolution imaging to determine whether the FFP is bound or Unbound.
In Section~\ref{sec:parallax}, we show that measuring the microlens parallax,
$\bpi_\e$, is extremely challenging for large-$\rho$ FFPs, although this
only presents fundamental difficulties for the Unbound among them.
In Section~\ref{sec:integrated}, we sketch the science that can be extracted
from measuring the FFP mass functions for each of the four categories
(Wide, Kuiper, Oort, Unbound).  We show how these can be extracted from
the {\it Roman} observations, augmented by late-time (e.g., 5--10 years later)
high-resolution observations, and possibly microlens-parallax observations.
In Section~\ref{sec:add-ben}, we show that our proposed changes to the
{\it Roman} observing strategy are beneficial to the remaining revolutionary
aspects of the original {\it Roman} program, namely low-mass bound planets
and wide-orbit planets.  We describe various other benefits of the change.
We also present
strategy options that represent more of a compromise with the
``old order'', although we do not advocate these.

\section{{What Is Killing the {\it Roman} Low-mass FFPs?}
\label{sec:killing}}

The origin of the drastic decline in detections from Earth-class to Mars-class
to Moon-class FFPs that is tabulated in Table~2 from \citet{johnson20}
is not what it may naively appear.  To understand this analytically, we
adopt their assumption of uniform surface brightness (no limb darkening).
We work in the limit $\rho\gg 1$, (or $\theta_\e\ll\theta_*$), i.e., the
regime of the lowest-mass detectable FFPs where the current observing
strategy is losing sensitivity,
and we assume that the blended light
is negligible.  The last assumption will be reviewed more closely for
various cases (see Sections~\ref{sec:role-gamma} and \ref{sec:solution})
but in general, is mostly relevant to
$M\la M_{\rm Mars}$ FFPs.
  
\begin{figure}
\plotone{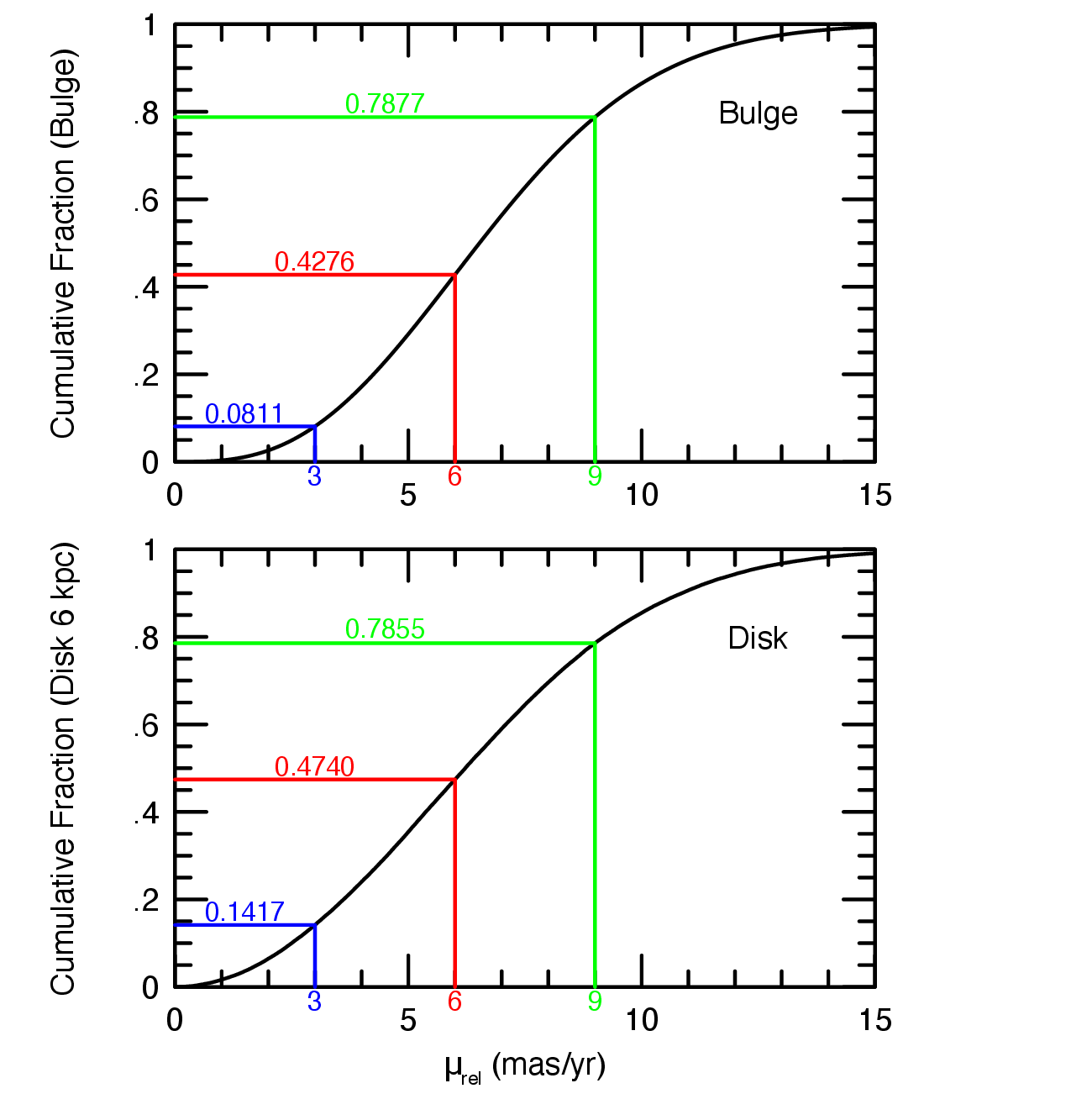}
\caption{Top Panel: Cumulative $\mu_\rel$ distribution based on the
  source and lens both being drawn from 2-dimensional isotropic
  Gaussians with $\sigma = 3\,\masyr$, appropriate for bulge lenses.
  At the currently adopted cadence ($\Gamma = 4\,{\rm hr}^{-1}$) and a
  ``typical'' value of $\theta_*=0.3\,\mu$as, proper motions
  $\mu_\rel\leq 3\,\masyr$ (blue) would be required to achieve 6
  magnified observations, which is the first requirement for an FFP
  detection.  This would capture only 19\% of the $\mu_\rel\leq 6\,\masyr$
  (red) range that would be available if the cadence were
  doubled to $\Gamma = 8\,{\rm hr}^{-1}$.  Bottom Panel:  Similar but
  for disk lenses: the corresponding fraction is 30\%.
}
\label{fig:cum-mu}
\end{figure}

Under these assumptions and limits, the magnification is given by
\begin{equation}
  A(u,\rho)-1 = {2\over\rho^2}\Theta(\rho-u);\qquad
  u = \sqrt{u_0^2 + \biggl({t-t_0\over t_\e}\biggr)^2}.
  \label{eqn:A-high-rho}
\end{equation}
Hence, the signal-to-noise ratio for a single exposure,
assuming photon-statistics, is
\begin{equation}
  {\rm S\over N} = {[A(u,\rho)-1]f_s\over\sqrt{A f_s}}\simeq
  {2\over\rho^2}\sqrt{f_s}\Theta(\rho-u)
  = 2\theta_\e^2{\sqrt{f_s}\over \theta_*^2}\Theta(\rho-u),
  \label{eqn:SN-high-rho}
\end{equation}
where $f_s$ is expressed in instrumental photon counts.
The expected number of magnified images is
\begin{equation}
  N_{\rm exp} = {2\Gamma \theta_*\over\mu_\rel}=
  3.0\biggl({\Gamma\over 4\,\rm hr^{-1}}\biggr)
  \biggl({\theta_*\over 0.3\,\mu\rm as}\biggr)^{}
  \biggl({\mu_\rel\over 6\,\masyr}\biggr)^{-1}
  \biggl({\beta\over 0.86}\biggr); \qquad \beta \equiv  \sqrt{1-z^2},
  \label{eqn:N-high-rho}
\end{equation}
where $z=u_0/\rho$ and $\Gamma$ is the survey cadence.
And therefore, the expected $\Delta\chi^2$ is given by
\begin{equation}
  \Delta\chi^2_{\rm exp} =
  N_{\rm exp} \biggl({\rm S\over N}\biggr)^2 = 
  8\theta_\e^4{\Gamma\over\mu_\rel}V_{H,\rm flux}\beta,
  \label{eqn:chi2-exp}
\end{equation}
where $V_{H,\rm flux}\equiv f_s/\theta_*^3$.  Equation~({\ref{eqn:chi2-exp})
 can then be evaluated,
\begin{equation}
  \Delta\chi^2_{\rm exp} =
  740\biggl({\theta_\e \over 0.1\,\mu\rm as}\biggr)^{4}
  \biggl({\mu_\rel \over 6\,\masyr}\biggr)^{-1}
  \biggl({\Gamma \over 4\,\rm hr^{-1}}\biggr)
  \biggl({V_{H,\rm flux} \over 1.85\,\times 10^5/\mu\rm as^3}\biggr)
  {\beta \over 0.86}.
  \label{eqn:chi2-exp2}
\end{equation}
We have expressed this evaluation in terms of the ``volume brightness''
\begin{equation}
  V_{H,\rm flux}\equiv {f_s\over \theta_*^3} =
  10^{-0.4\,A_H}{L_H/4\pi D_s^2\over (R_s/D_s)^3} = 10^{-0.4 A_H} D_s V_{H,\rm lum};
  \qquad V_{H,\rm lum}\equiv {L_H\over 4\pi R_s^3},
    \label{eqn:Vflux}
\end{equation}
because along the main sequence, $0.2\,M_\odot\la M\la 0.9\,M_\odot$,
$V_{H,\rm lum}$ is approximately constant.  We have written out the dependence
of the relation between $V_{H,\rm flux}$ and $V_{H,\rm lum}$ on the extinction
and distance for clarity.  However, in what follows, we will fix $D_s=8\,\kpc$
and the extinction $A_H=0.5$.  And we will treat $V_{H,\rm lum}$ as exactly
constant, so that $V_{H,\rm flux}$ is also exactly constant.  
Alternatively, because $\theta_\e^2=\kappa M \pi_\rel$,
Equation~(\ref{eqn:chi2-exp2}) can also be written as
\begin{equation}
  \Delta\chi^2_{\rm exp} =
  1100\biggl({M \over 0.01\,M_\oplus}\biggr)^{2}
  \biggl({\pi_\rel \over 50\,\mu\rm as}\biggr)^{2}
  \biggl({\mu_\rel \over 6\,\masyr}\biggr)^{-1}
  \biggl({\Gamma \over 4\,\rm hr^{-1}}\biggr)
  \biggl({V_{H,\rm flux} \over 1.85\,\times 10^5/\mu{\rm as}^3}\biggr)
  {\beta \over 0.86}.
  \label{eqn:chi2-exp3}
\end{equation}

We can now answer the question of why {\it Roman} has virtually no
sensitivity to $M=0.01\,M_\odot$ objects according to the \citet{johnson20}
simulations, as tabulated in their Table~2.  Clearly the answer
is not a lack of S/N: one just has to compare their $\Delta\chi^2>300$ criterion
to the normalization of Equation~(\ref{eqn:chi2-exp3}) at its fiducial
parameters.  Based on $\Delta\chi^2$ alone, such sub-Moons would be detectable
for all main-sequence sources $M_s>0.2\,M_\odot$, all proper motions
$\mu_\rel < 6\,\masyr$, and all relative parallaxes
$\pi_\rel\ga 25\,\mu$as.

The problem is rather located in Equation~(\ref{eqn:N-high-rho}): at the
fiducial parameters, there will be only three non-zero measurements, whereas
\citet{johnson20} require at least six 3-$\sigma$ measurements.
This requirement is reasonable.  While we do not presently know the exact
number that will be required, it would
certainly be impossible to interpret a detection with only 3 measurements,
and quite difficult with 4.  Until the actual data quality can be assessed,
a minimum of 6 points appears prudent.
So the limiting factor for detecting FFPs is the number of
magnified points rather than their individual (or combined) S/N.

In order to determine from Equation~(\ref{eqn:N-high-rho}) which FFP will
be detected and which will not,
one must first investigate the roles of each of the four scaling parameters:
$(\Gamma,\theta_*,\mu_\rel,z)$.  We examine these sequentially in reverse order.
There is almost no room for improvement in the $z$ scaling, which in any
case is a random and purely geometric factor.

\subsection{{Role of $\mu_\rel$}
\label{sec:role-murel}}

Regarding $\mu_\rel$, Equation~(\ref{eqn:N-high-rho}) is scaled to a
typical value for microlensing fields, the prospective {\it Roman}
fields in particular.  One can, in principle consider only the slower
events, e.g., $\mu_\rel=3\,\masyr$, at which the events will
have 6 points (keeping the other parameters the same).

To gain analytic understanding, we consider the ideal case of bulge-bulge
lensing with the distributions of the sources and lenses each characterized
by a 2-dimensional isotropic Gaussian with $\sigma=3\,\masyr$.
Keeping in mind that the event rate is proportional to $\mu_\rel$, the
distribution of event proper motions is
$f(\mu)d\mu \propto \mu^2 d\mu \exp(-\mu^2/4\sigma^2)$, or to simplify the
algebra,
\begin{equation}
  \label{eqn:gofx}
  g(x)dx ={2\over\sqrt{\pi}}x^{1/2}e^{-x} dx; \qquad
  x\equiv \biggl({\mu\over 2\sigma}\biggr)^2.
\end{equation}
    It is immediately clear from this formula
that only a fraction
\begin{equation}
 p= \int_0^{1/4}dx\,g(x)\la {2\over 3}\,{2\over\sqrt{\pi}}
  \biggl({1\over 4}\biggr)^{3/2} \sim 0.09
  \label{eqn:int-gofx}
\end{equation}
of the
distribution will have $\mu_\rel<\sigma = 3\,\masyr$.  Figure~\ref{fig:cum-mu}
shows the full cumulative distribution in the upper panel.

We also consider the case of disk lenses (and bulge sources).  For
this purpose, we adopt $D_l=6\,\kpc$, $D_s= 8.2\,\kpc$, a disk rotation
speed of $v_\rot=240\,\kms$, an asymmetric drift of $v_{\rm asym} =
25\,\kms$, bulge proper motion dispersions $(3.2,2.8)\,\masyr$ in the
$(l,b)$ directions, disk velocity dispersions of $(64,41)\kms$, and
solar motion $(+12,+7)\kms$.
The lower panel of Figure~\ref{fig:cum-mu} shows the resulting
cumulative distribution for the disk.

We return to a discussion of this figure in Section~\ref{sec:role-gamma}.

\subsection{{Role of $\theta_*$}
\label{sec:role-thetastar}}

From the form of Equation~(\ref{eqn:N-high-rho}), it is clear that
by doubling $\theta_*$ one can also double $N_{\rm exp}$, which would
bring it to the required six 3-$\sigma$ points for
the fiducial parameters of this equation.  Of course, the cost of
relying on such bigger (solar-type) source stars is that they are
much rarer than the early M-dwarfs that are used to scale the relation.
Indeed, the main point of conducting microlensing from space and in the
infrared is to access these much more numerous stars.

We can understand the role of $\theta_*$ as a continuous variable as
follows.  Because we are treating $D_s$ as constant, and because
$R\simeq R_\odot (M/M_\odot)$ on the main sequence (below the
turnoff), we have $\theta_*\propto M_s$.  The cross section for events
in the $\rho\gg 1$ regime that we are investigating scales as
$d\Gamma_{\rm events}/d\log M_s\propto
\theta_*(M_s)(M_sdN/dM_s)\propto M_s^2 dN/dM_s$.  That is,
$d\Gamma_{\rm events}/d\log M_s\propto (M_s/M_\odot)^{\alpha + 2}$ for
the case that the mass function (of sources) is described by a power
law, $\alpha$.

Based on {\it Hubble Space Telescope (HST)} optical counts of bulge stars by
\citet{sweeps}, we adopt a broken power law, with break point
$M_{\rm br} = 0.56\,M_\odot$, and powers $\alpha_{\rm high} = -2.41$ and
$\alpha_{\rm low} = -1.25$, respectively above and below the break.
Thus, $(\alpha+2)$ changes sign ($-0.41$ to $+0.75$) at the break,
implying that on a log-$M$ plot, there is a peak at early M-dwarfs.
In Figure~\ref{fig:thetastar}, we express this rate in terms of
$\theta_*$, by first employing the above approximations, i.e.,
$\theta_* = R_s/D_s = R_\odot/D_s (M_s/M_\odot) = 0.58\,\mu{\rm as}(M_s/M_\odot)$.
We express this relation in terms of $\theta_*$ (rather than $\log\theta_*$)
because it is more familiar.

We have extended the plot to the full main sequence ($1\ga M/M_\odot \ga 0.1$)
for clarity, noting that while the above S/N relation only applies in a
more limited range ($0.9\ga M/M_\odot \ga 0.2$), this relation does not
play a direct role in the current discussion.  Figure~\ref{fig:thetastar}
singles out the cumulative distribution up to fiducial value of
$\theta_*=0.3\,\mu$as, as well as for two other values, whose significance
will be made clear in Section~\ref{sec:role-gamma}.

\subsection{{Role of $\Gamma$}
\label{sec:role-gamma}}

The only other scaling variable that can be changed in
Equation~(\ref{eqn:N-high-rho}) is the observing
cadence, which is currently being set for {\it Roman} at the indicated
scaling value, $\Gamma=4\,{\rm hr}^{-1}$.

Of course, it requires no special insight to realize that by doubling
$\Gamma$, one also doubles $N_{\rm exp}$, albeit at the cost of
halving the number of fields (and so, the total area) that can be
observed.  However, making use of the results in
Sections~\ref{sec:role-murel} and \ref{sec:role-thetastar}, we are now
in a position to understand the impact of such doubling on the rate of
FFP detection in the large-$\rho$ (i.e., low-$M$) limit.

From Equation~(\ref{eqn:N-high-rho}), we see that one can, in
principle, reach the same adopted threshold for FFP detections,
$N= 6$, by either halving
$\mu_\rel$ or doubling $\Gamma$.  However, from
Figure~\ref{fig:cum-mu}, we see that by doing the first, we cut the
fraction of the cumulative $\mu_\rel$ distribution for bulge lenses
from 43\% (red) to 8\% (blue), i.e., by a factor of 5.3.  While doubling
$\Gamma$ comes at the cost of halving the number fields, there is
still an overall net increase in large-$\rho$
FFP detections of a factor 2.6.  The corresponding numbers for the
disk cumulative distribution are 47\% (red), 14\% (blue), and factors 3.3
and 1.7.

Motivated by this insight, one might consider increasing $\Gamma$ by a
further factor of 1.5 to $\Gamma=12\,{\rm hr}^{-1}$, which would allow
one to capture 79\% (green) of the bulge-lens $\mu_\rel$ distribution, i.e.,
a further increase by a factor 1.8.  Again, this would come
at a price of reducing the number of fields by a factor 2/3, implying
a net improvement of a factor 1.2.  This factor is quite minor, and such a
change would come at significant cost to other aspects of the
experiment.  A virtually identical argument applies to the disk-lens
proper-motion distribution.

Figure~\ref{fig:thetastar} allows us to make a similar evaluation for
the trade offs between changes in $\Gamma$ and $\theta_*$.  Comparing
the red and blue lines, one sees that restricting the mass (or luminosity,
or $\theta_*$) function to stars with $\theta_*>0.5\,\mu$as (which would
by itself not quite achieve the required doubling of $N_{\rm exp}$), would
reduce the available cumulative $\theta_*$ distribution function by almost
a factor of 5.  This is similar to the case for the bulge-lens $\mu_\rel$
distribution that was just discussed.

As in that case, we can also ask about the impact of a further
increase of $\Gamma$ by a factor of 1.5, which would drive the minimum
  source radius down to $\theta_*=0.2\,\mu$as.  This is shown in green.  The
nominal improvement is a factor of 1.56, which would be almost exactly
canceled by the loss of area due to higher $\Gamma$.  In fact the
range of ``improvement'' $0.3< \theta_*/\mu{\rm as}< 0.2$, is actually
pushing the FFP detections into a regime in which the assumptions
underlying the S/N calculation start to break down, mainly because
blending becomes a much more serious issue.  Hence, there would
actually be a net loss of FFP detections, even ignoring the negative
impact on other aspects of the experiment of such a further increase
in $\Gamma$.

If the {\it Roman} cadence remains at $\Gamma = 4\,{\rm hr}^{-1}$, as
derived by optimizing the bound-planet detections, then every
large-$\rho$ event that is selected according to the criterion of six
3-$\sigma$ points will have a product $\beta\theta_*/\mu_\rel$ that
is at least twice as big as that given by the fiducial parameters given
in Equation~(\ref{eqn:N-high-rho}).
That is, the prefactor in this equation is 3.0, so the product of
the remaining factors must be $\geq 2$ to achieve 6 points.
Ignoring the narrow range available from the final term $(\beta)$,
this implies some combination higher $\theta_*$ and
lower $\mu_\rel$.  To properly account for this, we should allow
$\mu_\rel$ and $\theta_*$ to vary simultaneously, rather than holding
the other fixed as in Figures~\ref{fig:cum-mu} and \ref{fig:thetastar}.
We therefore find the cumulative distribution of the product of the
$\mu_\rel$ and $\theta_*$ factors from Equation~(\ref{eqn:N-high-rho}).
\begin{equation}
  \zeta_{\mu,\theta_*} = 
    \biggl({\mu_\rel\over 6\,\masyr}\biggr)^{-1}
    \biggl({\theta_*\over 0.3\,\mu\rm as}\biggr)^{}
  \label{eqn:factor}
\end{equation}
Because the prefactor in Equation~(\ref{eqn:N-high-rho}) is 3.0, while
the detection criterion is $N=6$, $\zeta>2$ is required under the
present {\it Roman} strategy, but $\zeta>1$ would suffice for
$\Gamma = 8\,{\rm hr}^{-1}$.  To evaluate the cumulative distributions
for the bulge (black) and disk (magenta) cases, we draw from the range
$0.58\,\mu{\rm as} > \theta_* >0.25\,\mu{\rm as}$, as well as from the
full proper-motion distribution. 

The blue and red lines in Figure~\ref{fig:comb} highlight the
cumulative distributions for the cases of $\Gamma = 4\,{\rm hr}^{-1}$
($\zeta=2$) and $\Gamma = 8\,{\rm hr}^{-1}$ ($\zeta=1$), respectively.
The ratios of the two are 3.63 and 2.74 for the bulge and disk
respectively.  For FSPL events, bulge lenses are 2.5 times more
frequent than disk lenses (Figure 9 of \citealt{gould22}).  Weighting
the two ratios by this factor, we obtain a net improvement of a factor
3.38.

As mentioned above, this improvement must
be divided by 2 because there will be half as many fields.

Finally, there will be some additional detections because $\Delta\chi^2$
will double, which will push some very-low-mass FFPs above the threshold,
e.g., $\Delta\chi^2>300$, as adopted by \citet{johnson20}.  For example,
Equation~(\ref{eqn:chi2-exp2}) in its current form predicts
$\Delta\chi^2 = 150$, for $\theta_\e =0.067\,\mu$as.  If $\Gamma$ were doubled,
then $\Delta\chi^2 = 300$, and it would cross the threshold of detection.
A lens of mass $M=M_{\rm Pluto}$ could then be detected provided that
$\pi_\rel > \theta_\e^2/\kappa M_{\rm Pluto} = 84\,\mu$as, which corresponds to
$D_l\la 4.8\,\kpc$.  Hence, depending on whether Plutos are common
(about which we presently have only the barest indication from our own
Solar System), there could be many additional detections of FFPs from this
class.

We adopt a net improvement of a factor 2 in low-mass FFP detections.

One might also consider other cadences than the two shown in
Figure~\ref{fig:comb}.  To avoid cluttering this figure, we present these
results in tabular form in Table~\ref{tab:zeta}.  The final column in this
table is a figure of merit, which takes account of both the added FFP
detections due to higher $\Gamma$ and the reduced area.  However,
it does not take account of the extra (or reduced) FFP detections due to
higher (or lower) $\Gamma$.

\section{{Can $\theta_\e$ Be Measured for Large-$\rho$ FFPs?}
\label{sec:thetae}}

The Einstein radius, $\theta_\e$, is a crucial parameter for
understanding the FFP mass distribution. In particular, if the
microlens parallax $\pi_\e$ is also measured, then the mass is
directly given by $M=\theta_\e/\kappa\pi_\e$.  But even if $\pi_\e$ is
not measured, so that $\theta_\e$ remains a degenerate combination of
two unknowns ($\theta_\e=\sqrt{\kappa M\pi_\rel}$), it is still one step
closer to the mass than the routinely measured $t_\e$, which is a
combination of three unknowns ($t_\e=\sqrt{\kappa M\pi_\rel}/\mu_\rel$).

On the surface, it would appear that there are serious challenges
for the measurement of $\theta_\e$ for large-$\rho$ FFPs.  We argue in
this section that, on the contrary, for the great majority large-$\rho$
FFPs, $\theta_\e$ will be measured with sufficient accuracy to achieve the
main scientific goals.  We first outline the apparent challenges
and then describe how they can be addressed.

\subsection{{Challenges}
  \label{sec:challenges}}

The usual method to measure $\theta_\e$ is to measure $\rho$ and then to
determine $\theta_*$ using the method of \citet{ob03262}.  In this
method, one measures source flux and color from fitting the light curve,
measures its offset from the clump in these variables, and then uses
tabulated color/surface-brightness relations to determine $\theta_*$.
Finally, one calculates $\theta_\e = \theta_*/\rho$.

The challenges arise because each of these steps is, individually,
either difficult or impossible for {\it Roman} large-$\rho$ FFPs.
Hence, carrying out all of them would appear hopeless.

The first problem is that very few, if any, {\it Roman} large-$\rho$
FFPs will have a color measurement.  These events have a total
duration $\Delta t_{\rm event} \leq 2t_* = 2\theta_*/\mu_\rel =
0.9\,{\rm hr}(\theta_*/0.3\,\mu{\rm as})/(\mu_\rel/6\,\masyr)$.  So,
first, because the alternate-band observations are taken only twice
per day, the chance is small that these will occur during the time that the
source is magnified.  This issue is well recognized.

However, what seems to be less recognized is that if the second-band
observations are taken during the event, they will, in the overwhelming
majority of cases, prevent the light curve from being properly monitored
in the primary band.  This is because the secondary-band exposures are
much longer, so that to cycle through all the targeted fields requires
about 50 min.  Hence, the main impact of secondary-band observations
on FFPs will not be to measure their colors but to prevent the detection
of about 8\% of otherwise detectable events.

Second, as mentioned in Section~\ref{sec:two-tactics},
\citet{johnson22} show that these events display a strong degeneracy
between $f_s$ and $\rho$, meaning that in most cases, neither can be
measured separately.  Rather, what is measured is the parameter
combination $f_s/\rho^2$.  In particular, in the approximation of no limb
darkening, the excess flux as the lens is transiting the source is
just $\Delta f = 2f_s/\rho^2 $. In discussing this, \citet{johnson22}
point back to the fact that \citet{ob190551} had already shown that
this degeneracy is actually the key to measuring $\theta_\e$, provided
that the source color (hence surface brightness, $S_H$) is known.  Then
we can write $\Delta f = (2/\rho^2)\times (\pi\theta_*^2 S_H) =
2\pi\theta_\e^2 S_H$.  That is, $\theta_E^2 = \Delta f/2\pi S_H$.  Because
$\Delta f$ and $S_H$ are empirically determined quantities, $\theta_\e$
can be robustly measured even if $f_s$ and $\rho$ are not separately
measured.

The problem is that, as noted by \citet{johnson22}, {\it Roman} will yield
very few color measurements for large-$\rho$ FFPs.  Indeed, we should say
``essentially zero''.

\subsection{{Solution}
  \label{sec:solution}}

The solution to this seemingly intractable problem of measuring $\theta_\e$
(as opposed to either of the above two problems, considered individually)
comes in three
parts.  First, $\theta_\e$ measurements will span two decades, i.e.,
$0.1\,\mu{\rm as}\la \theta_\e \la 10\,\mu{\rm as}$.  Hence, we can
easily tolerate 10\% (0.04 dex) errors in typical individual
$\theta_\e$ measurements and even several tens of percent in some
subset of cases.  Second, one can estimate the surface brightness of the
source to within 20\% if its $H$-band luminosity is known exactly.
Third, errors in the inferred $\theta_\e$ scale only as the sixth-root
of errors in the luminosity.  In brief, adequate estimates of surface brightness
can be made without the customary source-color measurement.

For stars on or near the zero-age main sequence (ZAMS), their mass-radius
and mass-$L_H$ relations are determined by their
chemical composition.  Together, these algebraically predict the surface
brightness, $S_H$, as a function of $L_H$.  From isochrone models,
we know that in the $H$-band, the rms scatter in this relation is less
than 20\% over the range of bulge metallicities\footnote{
At fixed luminosity on the main-sequence, surface brightness $S_H$
only varies by $\la 10\%$ for metallicities [Fe/H]$\geq 0$.  It is higher
by $\sim 15\%$ for [Fe/H]$=-0.5$, and somewhat higher than that at
yet lower [Fe/H].  Thus, considering the distribution of metallicities
of microlensed bulge stars as measured by \citet{bensby17}, we find that the rms
error made by adopting the [Fe/H]$=0$ surface brightness, would be
$\sigma(\ln S_H)=10.3\%$.  However, a more precise evaluation, which
would account for $\alpha/$Fe variation, should be undertaken before
applying this method.}.  Because
$\theta_\e = \sqrt{\Delta f/2\pi S_H}$, such 20\% errors in $S_H$ lead
to only 10\% errors in $\theta_\e$.

This reasoning does break down for source stars $M_s\ga 0.9\,M_\odot$,
corresponding to $\theta_*\ga 0.53\,\mu$as in
Figure~\ref{fig:thetastar}, because these stars have moved off the
ZAMS by different amounts depending on their age.  However, first, we
can see from Figure~\ref{fig:thetastar}
that these stars account for a small fraction
of large-$\rho$ FFP events.  Second, the stars themselves are both
bright and sparse (in {\it Roman} data), so they will be only weakly
blended in the great majority of cases (unless the FFP has a bright
host, which can be determined from late-time high-resolution data).
Therefore, their color (hence, surface brightness) can be well
estimated from their well-measured color at baseline.

Finally, the luminosity $L_H$ can be estimated using
the relation $L_H = f_s \times 4\pi D_s^2 \times 10^{0.4 A_H}$,
where $f_s$ is estimated from the baseline flux,
$D_s$ is estimated from the mean distance of bulge sources in the
direction of the event, and $A_H$ is measured in the standard way from
field-star photometry. Clearly, then, this estimate of $L_H$ can
only be in error due to some combination of errors in $f_s$, $D_s$, and
$A_H$.

Before assessing these three error sources, we note that because
$V_{H,\rm lum}=L_H/4\pi R_s^3$ is approximately invariant over the relevant
range, $0.9\,M_\odot > M_s > 0.2\,M_\odot$, we have
$R_s\propto L_H^{1/3}$ and so, $S_H \propto L_H/R_s^2 \propto L_H^{1/3}$.
Hence, $\theta_\e^2\propto 10^{0.4\,A_H}\Delta f/S_H \propto
10^{0.4\,A_H}L_H^{-1/3}$, i.e., $\theta_\e \propto 10^{0.2\,A_H}L_H^{-1/6}$.
Assuming for the moment (as is almost always the case) that
$A_H$ is well measured, this implies that errors in the luminosity
estimate propagate to the $\theta_\e$ measurement only as the sixth-root.

Now, let us consider the three sources of error in $L_H$.  First, if
there is no parallax estimate for the source, then the rms error in
the source distance $D_s$ (due to the depth of the bulge) is about
10\%, leading to a 20\% error in $L_H$ and therefore a 3\% error in
$\theta_\e$, which is negligible in the current context.

Second, if the estimated $A_H$ is higher than the true one by $\Delta H$,
then $L_H$ will have been overestimated by a factor $10^{0.4\,\Delta H}$, and
therefore $\theta_\e$ will be overestimated by a factor
$10^{0.2\,\Delta A_H}\times (10^{0.4\,\Delta A_H})^{-1/6} = 10^{\Delta A_H/7.5}\sim
1 + 0.12\Delta A_H$.  Given that typical errors in $A_H$ are
$\sigma(A_H)\la 0.1$, this factor is also negligible.

Hence, the main issue is potential errors in $f_s$ due to blending
with some other star.  If the FFP is Unbound, then the only
possibilities are a companion to the source or an unrelated ambient star.
If it is bound, then there are two additional possibilities: the host
and/or a stellar companion to the host.

With the exception of the companion to the source, all of these
potential blends will be moving at a few $\masyr$ relative to the
source.  Hence, they all can be resolved and identified by taking
late-time high-resolution images using extremely large telescopes (ELTs).
In particular, the European Extremely Large Telescope (EELT)
will achieve 4 times better resolution than Keck,
i.e., 14 mas, just a few years after {\it Roman} launch.
Such late-time high-resolution images will be necessary in any case
in order to identify or rule out possible hosts of the FFP.  See
Section~\ref{sec:all-resolve}.

The main danger would therefore be
companions to the source.  With Keck resolution, these could be resolved
out for projected separations $a_\perp \ga 400\,\au$, while EELT could resolve
them down to $a_\perp \ga 100\,\au$.  Thus, about half of all binary-source
companions would escape detection regardless of effort.  Perhaps, half of
M-dwarfs have a companion, so about 1/4 of all Unbound detections would
have unresolved blended light from a source companion.

However, given the weak, $\theta_\e\propto L_H^{-1/6}$ scaling
relation, this would make very little difference if the companion were
fainter than the source, which is a substantial majority of cases.
For example, adopting the upper limit of this regime, i.e., an equal
brightness companion, $\theta_\e$ would be underestimated by a factor
$2^{-1/6}=0.89$, i.e., a 10\% error.  Of course, there would be cases
for which the binary companion was a few times brighter than the
source itself (and assuming that there were no clues to this in the
light curve), these might escape detection.  However, these would be
rare and, to take the relatively extreme example that the companion
was 1 mag brighter than the true source (yet still no clues to its
presence), the error would still only be a factor $3.5^{-1/6}= 0.81$,
which is tolerable for an occasional error, given the 2-decade range
over which $\theta_\e$ is being probed.

\section{{Two Issues Related to FFPs in Bound Orbits}
  \label{sec:bound}}

\subsection{{All FFP Candidates Require High-Resolution Imaging}
\label{sec:all-resolve}}

Even if there is no indication that the FFP has a host (such as a
disagreement between the positions of the event and the apparent
baseline object; or excess light superposed on the source in {\it
  Roman} images for cases that $f_s$ is well measured), it is still
necessary to search for possible hosts.  That is, even if the baseline
object appears to be consistent with what is derived from the event
about the source flux, there still could be a several times fainter
object that is superposed, which is either the host or the true source
(with the baseline object being dominated by the host).

The choice of the earliest time for making these observations would be
greatly facilitated by a measurement of $\mu_\rel$.  These will
usually, but not always (see Section~\ref{sec:mu-rel-difficult}), be
available for FSPL events, but they will never be available for PSPL
events\footnote{In some cases, however, there will be useful lower limits
on $\mu_\rel$ from upper limits on $\rho$.  While these should be derived
by fitting the actual light curve to a 5-parameter $(t_0,u_0,t_\e,\rho,f_s)$
model, a useful rule of thumb is: $\rho_\max =u_0\sqrt{u_0^2 +4}$, which can
be derived by equating the peak PSPL magnification,
$A_{\max,\rm PSPL}=(u_0^2 + 2)/u_0\sqrt{u_0^2 +4}$, with the peak FSPL
magnification (under the assumption that the lens transits the center
of the source): $A_{\max,\rm FSPL}=\sqrt{1 + 4/\rho^2}$.  Then,
$\mu_\rel > (\theta_*/\rho_\max)/t_*$.}.
Thus, the PSPL events will require some conservative guess
for when to take the first high-resolution followup observation.  In
principle, one might choose to forego PSPL events, or give them lower
priority.  However, PSPL events may provide the main window for
studying the higher-mass portion of the FFP mass function.  See
Sections~\ref{sec:without-par} and \ref{sec:parsat-pspl}.  An
intermediate approach would be to focus first on the FSPL FFPs ordered
by highest proper motion, and then start the PSPL FFPs.

\subsubsection{{Event-based $\mu_\rel$ Measurement Can Be Difficult in the
  Large-$\rho$ Limit}
  \label{sec:mu-rel-difficult}}

For the large-$\rho$ FFPs, which are of particular interest because they
probe the lowest masses, accurate proper-motion measurements can be challenging.
As discussed by \citet{johnson22}, in addition
to the degeneracy between $\rho$ and $f_s$, there is also a degeneracy between
$z$ and $t_*\equiv \rho t_\e$.  The quantity that is robustly measured from
the light curve is the time that source is significantly magnified, which
in the limit of $\rho\rightarrow\infty$, is just
$\Delta t_{\rm chord} = 2\beta t_*$.
Hence, expressed in terms of robustly
measured empirical quantities,
$\mu_\rel = \theta_*/t_* = 2\beta\theta_*/\Delta t_{\rm chord}$.
Even assuming that
$\theta_*$ has been accurately estimated from the source flux
(and possibly color), the estimate of $\mu_\rel$ is still directly
proportional to $\beta$.  The information on $\beta$ comes from the
amount of time required for the Einstein diameter to cross the limb
of the source ($\Delta t_{\rm limb}= 2\,t_\e/\beta$ to a good approximation
for $\beta\gg 0$)
compared to the time it spends transiting the
source ($\Delta t_{\rm chord} = 2\beta t_*$).  That
is\footnote{
There is a tight analogy between this equation (and indeed the whole
FSPL $\rho\gg 1$ microlensing formalism), and the formalism of
transiting planets.  However, there are three differences.  First, of
course, microlensing generates flux bumps while transits generate flux
dips.  Second, the size of these bumps/dips differs by a factor two,
i.e., $\Delta F = +2\pi \theta_\e^2 S $ for microlensing and $\Delta F
= -1\pi \theta_{\rm planet}^2 S $ for transits, where $S$ is the
source surface brightness and
$\theta_{\rm planet}\equiv r_{\rm  planet}/D_{\rm planet}$.
And third, more subtly, while the
transit deficit arises from an opaque body, effectively an integral
over a 2-dimensional $\Theta$ function of radius $\theta_{\rm
  planet}$, the microlensing excess arises from a smooth, though
relatively compact excess magnification function,
$\Delta A_{\mu\rm Lens} = [(u^2 + 2)/u\sqrt{u^2+4}]-1$, with effective radius
$\sqrt{2}\theta_\e$, where $u\equiv \theta/\theta_\e$.  The difference in
compactness of these two
examples can be quantified in terms of a dimensionless concentration parameter
$C \equiv \pi\langle\theta\rangle^2/h_0$ where
$\langle\theta\rangle\equiv h_1/h_0$ and
$h_n\equiv \int d\Omega\,\Delta A\theta^n$.  For microlensing
$C_{\mu\rm Lens} = 8/9$.  For transits, with
$\Delta A = -\Theta(\theta_{\rm planet} - \theta)$, one finds
$C_{\rm transit} = -4/9$.  Hence, microlensing produces somewhat less
distinct features than transits as the planet transits the limb of the source.
},
$\beta = \sqrt{\Delta t_{\rm chord}/\Delta t_{\rm limb}}/\rho$, or
\begin{equation}
  \mu_\rel = 2\sqrt{{\theta_\e\over \Delta t_{\rm chord}}\,
    {\theta_*\over \Delta t_{\rm limb}}}.
  \label{eqn:beta-degen}
\end{equation}

Both quantities in the first ratio in Equation~(\ref{eqn:beta-degen})
can be robustly measured.  In many cases, the numerator of the second
ratio ($\theta_*$) can be well determined.  However, measurement of
the denominator ($\Delta t_{\rm limb}$) depends on good measurements
during the  brief intervals of the limb crossings, which may be difficult,
particularly for $\rho\gg 1$.  Hence $\mu_\rel$ measurements are likely
to be much more robust for events in the regime $\rho\la 2$ than in
the large-$\rho$ limit.

However, if $\theta_*$ can be measured, then even if $\mu_\rel$ cannot
be measured, one still obtains an upper limit
$\mu_\rel = 2\beta\theta_*/\Delta t_{\rm chord}\leq 2\theta_*/\Delta t_{\rm chord}$
because $\beta\leq 1$.  Moreover, in a substantial majority of cases, $\mu_\rel$
will actually be near this limit because for 60\% of random trajectories,
$\beta> 0.8$, while for large-$\rho$ events, the $\sim 13\%$ of trajectories
with $\beta<0.5$ are unlikely to yield viable events due to the paucity of
magnified points.  While this soft limit is likely to play little role in
the scientific analysis of these events, it can play a practical role
in deciding when to take late-time AO observations.

\subsection{{Possible Reduction of the $\Delta\chi^2$ Threshold for Kuiper FFPs}
\label{sec:kuiper-special}}

We have adopted a threshold $\Delta\chi^2>300$ for FFP detection following
\citet{johnson20}, which is substantially larger than the $\Delta\chi^2>60$
threshold for short planetary perturbations on otherwise 1L1S events, which
thereby transform them into double-lens single source (2L1S) events.
While both numbers may change in the face of real data, it is certainly
correct that the first should be much larger than the second.

There are two reasons for this.  The main one is that the effective number
of trials is vastly greater for the FFP search, which probes $\sim 2\times 10^8$
sources over six distinct seasons, compared to the 2L1S search, which probes
$\sim 10^5$ microlensing events, each basically contained in one season.
This is a ratio of $\sim 10^4$.  Secondarily, the FFPs are described by
5 parameters, whereas the 2L1S perturbations require only 4 additional
parameters because the source flux $f_s$ is already known from the main event.

However, a specific search for Kuiper FFPs would be triggered by the
presence of a star that is brighter than the apparent microlensed
source by at least 1 mag and lying within 1 {\it Roman} pixel of it
(but clearly offset from it).  Of all possible field stars that are the
apparent location of a microlensing event that must
be considered for such a search, $\la 1\%$ will have a neighboring
star that will generate a false positive by meeting these conditions.
This is due to the low surface density of such field stars, and the
small fraction of binary companions in this parameter range.  Hence,
rather than facing a factor $10^4$ more trials, there would only be a
factor $\la 10^2$.  Therefore, the $\Delta\chi^2$ threshold could
perhaps be reduced by a factor 2/3 for such Kuiper candidates without
burdening the search with too many false positives, thereby increasing
the sensitivity to very low-mass Kuiper FFPs.

\section{{Microlens Parallax Measurements for FFPs}
  \label{sec:parallax}}

As we discuss in Section~\ref{sec:integrated}},
the microlens parallax, $\bpi_\e$,
\begin{equation}
  \bpi_\e \equiv \pi_\e {\bmu_\rel\over \mu_\rel}; \qquad
  \pi_\e \equiv {\pi_\rel\over\theta_\e} = \sqrt{\pi_\rel\over\kappa M},
  \label{eqn:bpie}
\end{equation}
has a wide variety of applications for FFPs, assuming that it can be measured.
These go far beyond the most widely recognized applications that, when
combined with a measurement of $\theta_\e$, the microlens parallax
immediately yields the lens mass $M=\theta_\e/\kappa\pi_\e$
and the lens-source relative parallax $\pi_\rel=\theta_\e\pi_\e$
\citep{gould92}, which then yields the lens distance
$D_l = \au/(\pi_s+ \pi_\rel)$ provided that $\pi_s$ is at least
approximately known.

In this section, we focus on determining the lens characteristics
for which $\bpi_\e$ is measurable.

\citet{refsdal66} originally advocated Earth-satellite parallaxes
based on a principle that is well-illustrated by Figure~1 of
\citet{gould94}.  This concept was extended to Earth-L2 parallaxes by
\citet{gould03} specifically as a method to obtain microlens
parallaxes for terrestrial-mass objects.  The choice of the
Earth-satellite projected baseline $D_\perp$ is relevant because if the
satellite lies too far inside the Einstein ring projected on the
observer plane, ${\tilde r}_\e \equiv \au/\pi_\e$, i.e.,
$D_\perp\ll {\tilde r}_\e$, then the Earth and satellite light curves
will be too similar to measure the parallax effect, while if it lies
too far outside, $D_\perp\gg{\tilde r}_\e$, there will be no
microlensing signal at one of the two observatories.  Hence, for a
given targeted $\pi_\e$, one should strive for
\begin{equation}
  0.05< \pi_\e {D_\perp\over\au} < 2;\qquad  ({\rm Refsdal\ Limit}).
  \label{eqn:refsdal-limit}
\end{equation}
  
For Earth-mass lenses, and adopting
$D_\perp=0.01\,\au$ for L2 at quadrature (the mid point of {\it Roman}
observations), this implies an accessible range of
$\pi_\rel=\pi_\e^2\kappa M_\oplus$ of $0.6\,\mu{\rm as}\la \pi_\rel\la
1000\,\mu{\rm as}$, which encompasses nearly the full range of relevant lens
distances.  This is the reason that L2 parallaxes are ideal for
terrestrial planets.
More generally, the two red lines in Figure~\ref{fig:par} show these
boundaries on the $(\log M,\log\pi_\rel)$ plane.

Figure~\ref{fig:par} shows a second relation, the ``Paczy\'nski Limit''
(magenta), that further bounds the region of ``Earth + L2-Satellite''
parallax measurements, which is overall outlined in green.  This limit
is given by the inequality, $\rho \equiv \theta_*/\theta_\e <2$, i.e.,
\begin{equation}
  \theta_\e >{\theta_*\over 2};\qquad  ({\rm Paczynski\ Limit}).
  \label{eqn:pac-limit}
\end{equation}
For purposes of illustration, we have adopted $\theta_*=0.3\,\mu$as,
which is the most common class of source star that will enter in FFP
measurements.  We note that both Equations~(\ref{eqn:refsdal-limit})
and (\ref{eqn:pac-limit}) are somewhat soft and depend on the quality of
the data and the geometry of the event.  Nevertheless, given that the diagram
spans several decades in each directions, this softness is of relatively
small importance.

\subsection{{Parallax Measurements In the Large-$\rho$ Regime Are Difficult}
\label{sec:parallax-difficult}}

The origin of the Paczy\'nski limit is that the measurement of two of
the Paczy\'nski parameters ($u_0,t_\e$) are difficult unless
Equation~(\ref{eqn:pac-limit}) is satisfied.  These parameters enter
into the equation that describes the parallax measurement from two
observatories (see Figure~1 of \citealt{gould94}),
\begin{equation}
  \bpi_\e = (\pi_{\e,\parallel},\pi_{\e,\perp}) = {\au\over D_\perp}
  \biggl({{\Delta t_0\over t_\e},\Delta u_0}\biggr);\quad
  \Delta t_0 = t_{0,2} - t_{0,1}; \quad\Delta u_0 = u_{0,2} - u_{0,1},
  \label{eqn:par-measure}
\end{equation}
where  $(t_0,u_0)_{1,2}$ are the parameters measured for each observatory and
the two components of $\bpi_\e$ are, respectively, parallel and perpendicular
to the vector projected separation of the two observatories ${\bf D}_\perp$.
As already noted by \citet{refsdal66}, because $u_0$ is a signed quantity, but
only $|u_0|$ is usually measured, $\pi_{\e,\perp}$ is subject to a four-fold
degeneracy, including a two-fold degeneracy in
$|\pi_{\e,\perp}|_\pm = (\au/D_\perp)||u_{0,1}| \pm |u_{0,2}||$, which then
induces a two-fold degeneracy on the amplitude of $\bpi_\e$, i.e.,
\begin{equation}
\pi_{\e,\pm}=\sqrt{\pi_{\e,\parallel}^2 + \pi_{\e,\perp,\pm}^2}.
\label{eqn:piepm}
\end{equation}

In the context of large-$\rho$ microlensing, the determinations of the
two Paczy\'nski parameters,
$u_0=z/\rho$ and $t_\e = t_*/\rho$, depend directly on knowing $z$ and $t_*$,
even provided that $\rho$ is well-determined.

However, as demonstrated by \citet{johnson22}, there can be a strong degeneracy
between the source-star impact parameter, $z$, and source
self-crossing time $t_*=\rho t_\e$.  As discussed in
Section~\ref{sec:mu-rel-difficult}, the robust observable is $\Delta
t_{\rm chord} = 2\beta t_*$, i.e., the duration of the well-magnified
portion of the light curve, where, again, $\beta\equiv\sqrt{1-z^2}$.

Now, if $z$ can be measured from either of the two observatories, then it
will also be known for the other.  This is because by measuring $z_1$, one
infers $t_{*,1} = \Delta t_{\rm chord,1}/2\beta_1$, where
$\Delta t_{\rm chord,1}$ is a direct observable and $z_1$ is (by hypothesis)
measured.  But $t_{*,2}= t_{*,1} =t_*$ is the same for both observatories,
so that $\beta_2 = \Delta t_{\rm chord,2}/2 t_*$ is also a combination
of well-determined quantities.

However, if the $z$-$t_*$ degeneracies remain severe for both
observatories, the parallax measurement will be severely compromised
and difficult to exploit.  When the event is observed
from two observatories, the peak times $t_0$ are not degenerate
with any other parameters, so one can robustly infer that the component
of $\bpi_\e$ along the Earth-satellite axis is given by
$\pi_{\e,\parallel} = (\au/D_\perp)\Delta t_0/t_\e
= (\au/D_\perp)(2\Delta t_0/\Delta t_{\rm chord,1})(\theta_\e/\theta_*)\beta_1$.
Then, because $\beta_1\leq 1$ and all the other quantities are measured,
this places an upper limit on $\pi_{\e,\parallel}$.  However, because
$\pi_\e = \sqrt{\pi_{\e,\parallel}^2 + \pi_{\e,\perp}^2}$ and $\pi_{\e,\perp}$
is likely to be poorly constrained due to the \citet{refsdal66} four-fold
degeneracy combined with the poorly determined values of $z$ (and so
$u_0 = z\theta_*/\theta_\e$), there will be neither an upper nor a lower limit
on $\pi_\e$ in most cases.

We conclude that parallax measurements are unlikely to provide much information
for FFPs with $\rho\gg 1$.  Nevertheless, parallax measurements can provide
very useful information for FFPs with $\rho\la 2$, as discussed by
\citet{zhu16}, \citet{gould21}, and \citet{ge22}
for Earth-L2 parallaxes, by \citet{euclid19}, \citet{ban20}, 
\citet{gould21} and \citet{bachelet22}
 for {\it Roman}-{\it Euclid} parallaxes,
and by \citet{yan22} for {\it CSST}-{\it Roman} parallaxes.
Hence, provided any of these programs are executed, they will also automatically
provide parallax information on large-$\rho$ events, which could prove
useful in some cases.

\section{{Integrated Approach Toward an FFP Mass Function}
  \label{sec:integrated}}

In this section, we sketch how an ensemble of FFP detections in
eight categories (FSPL,PSPL) $\times$ (Wide,Kuiper,Oort,Unbound)
can be combined to measure the FFP
mass function as a function of FFP dimensionless binding energy, i.e.,
$v_{\rm orb}^2/2$, where $v_{\rm orb}$ is the orbital velocity of the bound
planet.
For the four orbit categories just listed, these are
roughly $v_{\rm orb} = (7, 3, 1, 0)\,\kms$. 

In fact, the paths for incorporating members of these eight categories
into the mass-function determination differ substantially from one
another.  Moreover, they differ between the case that microlens
parallax measurements are made or not.  To help navigate this somewhat
complex discussion, we begin with an overview of main issues that
affect all cases.  Next, we give concrete illustrations of how the FFP
mass functions of the four orbit categories can provide insight into
planet formation and early evolution.  We then carry out separate
discussion for the FSPL FFPs and PSPL FFPs. 

\subsection{{Overview of Issues Related to All Eight Categories}
  \label{sec:overview}}

The overriding issue is to distinguish between Unbound FFPs and the
three categories of bound FFPs by identifying the hosts of the latter group.
If the host for a bound FFP cannot be identified, then it is, in fact,
unknown whether the FFP is bound or not.  And to the degree that this is common,
the derived Unbound sample will be contaminated with bound FFPs.  While such
ambiguities are inevitable at some level, if they are frequent, then
the scientific investigations that are sketched in Section~\ref{sec:origins}
will become difficult.

In general, it will be far easier to identify the hosts of Wide and
Kuiper FFPs (whether FSPL or PSPL) than Oort FFPs because the hosts
of the former will be projected very near on the sky to the location
of the event.  Hence, the chance of a random interloper being
projected at such a close separation is low.  However, for Oort FFPs,
the chance of random-interloper projections can be close to 100\%.
Hence, the main issue is how to distinguish Oort FFPs from Unbound
FFPs, by securely identifying the hosts of the former, in the face of
a confusing ensemble of candidate hosts.

We will show that if $\bpi_\e$ is measured, then it is possible to
identify the host of Oort objects up to a considerable separation,
i.e., well into the regime of many random-interloper candidate hosts.
This is true for both FSPL and PSPL FFPs but will be more robust
for the former.

In brief, with $\bpi_\e$ measurements, it will be possible to
systematically identify hosts for all bound classes of FSPL FFPs and
to measure the masses, distances and transverse velocities of
essentially all of these.  The same basically holds for PSPL FFPs, but
the identifications will be more difficult.  It will also be possible
to measure the masses, distances and transverse velocities of the FSPL
Unbound FFPs (modulo the \citealt{refsdal66} four-fold
degeneracy). However, without $\bpi_\e$ measurements, only Wide and
Kuiper FFPs will have mass measurements, some of these measurements
will be quite crude, and there will be no mass measurements for
Unbound FFPs.  Hence, the premium on obtaining $\bpi_\e$ measurements
is extremely high.

As we showed in Section~\ref{sec:parallax-difficult}, it will usually
not be possible to measure $\bpi_\e$ for FFPs in the $\rho\gg 1$
regime, which contains the lowest-mass FFPs.  In order to illustrate the
size of this region of parameter space relative to the region for which
$\bpi_\e$ is measurable, we show (in Figure~\ref{fig:par})
the ``Detection Limit'' (in blue) of
$\theta_\e > 0.067\,\mu$as, by substituting $\Gamma=8\,{\rm hr}^{-1}$,
$\mu_\rel = 6\,\masyr$, and $z=0.5$ into Equation~(\ref{eqn:chi2-exp2})
and demanding $\Delta\chi^2_{\rm exp} > 300$.  Note that it lies 0.7 dex below
the Paczy\'nski Limit, corresponding to a factor $\simeq\sqrt{5}$ smaller
in $\theta_\e$, and so a
factor $\simeq 5$ smaller in planet mass at fixed $\pi_\rel$.

\subsection{{Possible Origins of the four classes of FFPs}
  \label{sec:origins}}

In this section we speculate on the origins of
the four categories of FFPs, i.e., Wide, Kuiper, Oort, and
Unbound.  The point is not to make predictions but to
illustrate how specific hypotheses on these origins can be tested
observationally by measuring the FFP mass functions of the four groups.

We begin with the Unbound objects.  Most likely, these
were ejected by planet-planet interactions (for
references on ejection mechanisms, see the relevant discussions in the recent
reviews by \citealt{zhudong} and \citealt{mroz23}).
If so, the ejecting planet
should have an escape velocity that is of order or greater than its
orbital velocity.  This applies in our own solar system to Jupiter, Saturn,
Uranus, and Neptune, but not to any of the terrestrial planets.  Moreover,
if these were at the position of Earth, it would still robustly apply to Jupiter
but only marginally to Saturn.  Hence, we conjecture that such objects
are mainly formed locally in the richest regions of the proto-planetary
disk, i.e., just beyond the snow line, where they are perturbed by gas giants.

Next we turn to Oort objects.  The process just described will inevitably
put some of the ejected objects in Oort-like orbits, but only a fraction of
order $\sim (v_{\rm Oort}/v_{\rm perturber})^2\sim 1\%$.  Thus, if this
hypothesis is correct,  Oort objects should have a similar mass function to
the Unbound objects, but be of order 100 times less numerous.
Alternatively, the Oort objects could have formed like our own Oort Cloud
is believed to, by repeated, pumping-type perturbations from relatively
massive planets far beyond the snow line.  In this case, the Oort objects
would have a different mass function, being in particular cut off at the high
end and perhaps different in form at lower masses as well.

If the Kuiper objects were the ultimate source of the Oort objects, as
just hypothesized, then they should have a similar mass function, but
in addition contain the perturbers, and they should also contain
additional objects that are below the perturber masses, but that are
deficient among Oort objects because they are too heavy to be pumped.

Finally, the Wide planets may simply be the members of the
ordinary bound planet population that happen to escape notice because
of geometry.  If so, they should have a similar mass function to these
bound planets.

Again, we emphasize that these are extreme-toy models and are in no 
sense meant to serve as predictions.  They are just presented to illustrate
the role of mass functions as probes of planet formation.

\subsection{{Measurement of the FFP Mass Function(s)}
  \label{sec:ffp-mf}}

We adopt the orientation that ELT observations can be made of all FFPs
after the source and host (or putative host) have separated enough to
resolve them.  And we further assume that, whenever necessary,
a second epoch can be taken to measure the host-source relative proper motion,
which (given the low orbital speeds of bound FFPs) will be essentially the
same as the lens-source (i.e., FFP-source) relative proper motion $\bmu_\rel$.
In principle, the number of such objects could be too large
for this to be a practical goal, but the logic outlined below can still
be applied to a well-selected subsample.

\subsubsection{{What Can Be Accomplished Without $\bpi_\e$ Measurements?}
  \label{sec:without-par}}

With or without microlens parallax measurements, the critical question
will always be whether (or to what extent) the hosts of the bound FFPs
can be identified.  If they can be identified, then first (obviously)
they can be identified as bound, in which case those that are actually
Unbound will be mixed together with the bound FFPs whose hosts have
simply not been identified as such.  Second, it will be possible to
derive mass, distance, and transverse-velocity measurements for these
bound FFPs.  The distance measurement will come from the photometric
distance estimate\footnote{Note that this implicitly assumes that the hosts
(or possibly stellar companions to the hosts)
of bound FFPs are luminous.  The practical implication of this assumption
is that FFPs that have dark hosts (such as white dwarfs and brown dwarfs)
that lack luminous stellar companions
will not be recognized as such, and therefore they will inevitably be
lumped together with Unbound FFPs in subsequent mass-function analyses.}
of the host (or possibly, a stellar companion to the host).
In some cases, it will be possible to
measure the trigonometric parallax of the host from the full time
series of {\it Roman} astrometric data.  However, these instances will
be extremely rare for the non-microlens-parallax case, so we discuss
this prospect within the context of microlensing parallax measurements, below.

The mass determination will come from combining measurements of the
$\bmu_\rel$ and $\pi_\rel$ (which very well approximate the
host-source relative proper motion and parallax).  The first of these
will come directly from two epochs of late-time ELT observations.
The second will come by combining the photometric host
distance with the fact that the source is in the bulge.  Then, using essentially
the method first proposed by \citet{refsdal64}, the FFP mass is given by
\begin{equation}
  M = {\theta_\e^2\over \kappa\pi_\rel} = {(\mu_\rel t_\e)^2\over \kappa\pi_\rel}, 
  \label{eqn:refsdal64}
\end{equation}
where either $\theta_\e$ or $t_\e$ is measured from the light-curve analysis.
The only difference relative to \citet{refsdal64} for the second ($t_\e$) case
is that he imagined that $\pi_\rel$ and $\mu_\rel$ would be measured for the
lens that generated the microlensing event, as opposed to a stellar companion
to the lens that was of order a million times more massive.

Note that both forms of Equation~(\ref{eqn:refsdal64}) are important.
For FSPL events, $\theta_\e$ can be measured even when $t_\e$ cannot,
in particular in the $\rho\gg 1$ limit.  Thus, masses can be derived
for these seemingly poorly measured objects (provided that their hosts
can be identified).  On the other hand, the PSPL FFPs, which dominate
the higher-mass FFPs (and so typically generate longer, better
characterized events), will often have well-measured $t_\e$ even
though they lack a $\theta_\e$ measurement.

Finally, the transverse velocity can be derived from the distance
and proper-motion measurements.

The main contributor to the error in the mass measurements will be the
accuracy of the $\pi_\rel$ estimate.  For disk lenses, combining the
knowledge that the source is in the bulge with the photometric lens
distance will
lead to a reasonably good ($\la 0.1\,$dex) estimates of
$\pi_\rel$ and hence, taking account of the $\sim 10\%$ errors in
$\theta_\e$, roughly 0.15 dex errors in
$M=\theta_\e^2/\kappa\pi_\rel$.  For bulge lenses, the $\pi_\rel$
errors will be more like a factor 2, so mass errors of 0.3 dex.

For the great majority of Wide and most Kuiper FFPs, it will be
possible to identify the host with reasonably good confidence based
primarily on proximity and supplemented by photometric estimates of
the candidate-host distance, and brightness.  The criterion will be
the probability that an unrelated star at the estimated distance could
be projected within the measured angular separation by chance.  For
example, stars whose photometric properties are consistent with them
being in the bulge (and brighter than $H_{\rm Vega}<21$) have a
surface density of a few per square arcsec.  Hence, if one appears
projected at $\Delta\theta=17\,\mas$ (corresponding to
$a_\perp=100\,\au$ at $D_l=6\,\kpc$), then the false alarm probability
(FAP) that it is unrelated to the event is $p\la 0.5\%$.  Hence, this
star would be judged as being associated with the event unless there
were another competing candidate (which would be very rare, i.e., just
the same $p=0.5\%$ of the time).  The possibility (far from
negligible) that the observed star was a companion to the source could
easily be ruled out by the ELT proper motion measurement.  Then the
star must be either the host or a stellar companion to the host.  The
latter possibility would have no impact on the distance estimate and
would be very unlikely to significantly affect either the mass or the
transverse-velocity estimates, although it would potentially affect
the estimate of $a_\perp$.

This argument could, by itself, be pushed to separations that are about
3 times larger, i.e., $a_\perp\sim 300\,\au$.
If the photometric distance estimate clearly excluded a bulge location
for the candidate, then the same method could be pushed yet further by
a factor of a few depending on the actual distance.  For FSPL FFPs, which
have scalar proper motion measurements, these could provide additional vetting
against false candidates.  

It is quite possible, in principle, that the overwhelming majority of
bound FFPs lie projected at $a_\perp$ values that are accessible to
this technique.  If so, this would likely become apparent from a rapid
fall-off of FFPs with $a_\perp$.  In this case, it would be reasonable
to assume that most FFPs that lacked hosts within the range accessible
to this technique were, in fact, Unbound FFPs, and in particular, that
there were very few Oort FFPs.

On the other hand, it is also possible that the bound FFPs extend to
larger separations than can be vetted by the techniques of this
section, and so require microlens-parallax measurements in order to
securely identify them.  Moreover, whether or not this improved
vetting of candidates proves necessary, $\bpi_\e$ measurements would
greatly improve the mass measurements for bound FFPs, and they would
provide the only possible direct mass measurements for Unbound FFPs.

\subsubsection{{Role of a Parallax Satellite for FSPL FFPs}
  \label{sec:parsat-fspl}}

The measurement of the microlens parallax vector, $\bpi_\e$, would provide
several key pieces of information with respect to measuring the mass
functions of the four different categories of FSPL FFPs.  First, of
course, it will essentially always yield the FFP mass,
$M=\theta_\e/\kappa \pi_\e$, and its lens-source relative parallax,
$\pi_\rel = \theta_\e\pi_\e$, because $\theta_\e$ will essentially
always be measured for FSPL FFPs that are accessible to $\bpi_\e$
measurements.  As discussed in
Section~\ref{sec:without-par}, mass measurements will be possible for
a large fraction of bound FFPs by identifying their hosts, even in the
absence of $\bpi_\e$ measurements.  However,
first, $\bpi_\e$-based mass measurements will be substantially more
accurate for disk FFPs and dramatically more accurate for bulge FFPs.
Second, without $\bpi_\e$ measurements, hosts cannot be identified at
very wide separations $a_\perp\ga 300\,\au$, so $\bpi_\e$-based mass
measurements are essential for these cases.  Third, for Unbound FFPs,
the only possible way to measure the mass is via microlens parallax.

However, of more fundamental importance is that $\bpi_\e$ measurements
will enable systematic vetting of candidate hosts and therefore allow
for robust identification of unique hosts, as well as robust
identification of FFPs that lack hosts (i.e., Unbound FFPs).  We say
``more fundamental'' because without host identification, one cannot
distinguish between classes of FFPs, in particular to determine which
are actually Unbound.  Moreover, for bound FFPs, the only way to resolve
the two-fold \citep{refsdal66} mass degeneracy that derives from the
$\bpi_\e$ measurement is by identifying the host.

Therefore, the remainder of this section is devoted
to the role of $\bpi_\e$ measurements in robust host identification of FSPL
FFPs.

The main technique for vetting candidates is to compare the observed
candidate-source vector proper motion derived from late-time ELT imaging with
the predicted vector lens-source proper motion
\begin{equation}
  \bmu_\rel = \mu_\rel{\bpi_\e\over \pi_\e},
  \label{eqn:pred-pm}
\end{equation}
where $\mu_\rel$ is the scalar proper motion that is derived from the
finite-source effects of the FSPL event.  A complementary, though less
discriminating, vetting method is to compare candidate-source $\pi_\rel$
(usually from photometric distance estimates from late-time imaging),
to the lens-source $\pi_\rel=\theta_\e\pi_\e$.

Recall that the \citet{refsdal66} four-fold
degeneracy consists of a two-fold degeneracy in $\pi_\e$, with each value
being impacted by a two-fold degeneracy in the proper-motion direction,
${\hat \bmu}_\rel$.  It is important to note that generally the two sets
of directional degeneracies do not overlap: see Figure~1 of \citet{gould94}.
The FSPL FFPs have measurements of the scalar $\mu_\rel$ from the light curve.
Bringing together all this information, a candidate host must be consistent
with one of four vector proper motions $\bmu_\rel$, all with the same amplitude
but with different directions.  And for each of these directions, there
is a definite value (one of two, see Equation~(\ref{eqn:piepm}))
of $\pi_\e$, implying a definite value of $\pi_\rel$.

Regarding Wide and Kuiper FFPs, we have already argued in
Section~\ref{sec:without-par} that excellent host identifications can
usually be made based primarily on proximity, together with other
information that does not require a $\bpi_\e$ measurement.
Nevertheless, making certain that the candidate's proper motion is
consistent with one of the four proper motions from
Equation~(\ref{eqn:pred-pm}) is a useful sanity check.  And in some
cases the proximity technique may result in multiple candidates, which
can be resolved based on the more stringent (vector) proper-motion requirement.
And in some further rare cases, the $\pi_\rel$ consistency check may
play a role.

Next, we consider Oort FFPs, which according to our schematic
characterization begin at $a_\perp \sim 300\,\au$, or
$\Delta\theta=50\,\mu$as for $D_l = 6\,\kpc$.  At this nominal
boundary, two aspects of the Kuiper situation would remain
qualitatively similar: the source and host would not be resolved in
{\it Roman} images at the time of the event and the FAP would still be
relatively small ($p\sim 5\%$) so that the chance of multiple random
interlopers at these separations would be small.  However, the FAP
would not be so small as to allow one to make a secure identification
based on proximity alone.  Nevertheless, with the amplitude of
$\bmu_\rel$ accurately predicted from the event, and its direction
predicted up to a four-fold degeneracy by the $\bpi_\e$ measurement,
it is very unlikely that the true host among the handful of candidates
(still, most likely only one), would not be identified.  Indeed, the
latter statement would remain qualitatively the same out to
$a_\perp=1000\,\au$ corresponding to $\Delta\theta=170\,\mas$, for
which $p\sim 0.5$.

However, at this and larger separations, the source and host would be
reasonably well resolved, and it would be possible, using {\it Roman}
data alone (that is, without waiting for late-time ELT observations)
to measure the candidate-source relative parallax and proper-motion,
$(\pi_\rel,\bmu_\rel)_{c-s}$ and thus to ask whether they were both consistent
with the values of these quantities that were derived from the event:
$(\pi_\rel,\bmu_\rel)_{l-s}=(\theta_\e\pi_\e,[\theta_\e/t_\e][\bpi_\e/\pi_\e])$.
Scaling from Figure~1 of \citet{gould15}, the individual-epoch astrometric
precision would be (ignoring any extra noise due to blending),
\begin{equation}
  \sigma_{\rm ast} = 1.3\,\mas 10^{0.2(H_{\rm Vega}-21)}
  \label{eqn:ast-prec}
\end{equation}
Because the $N=8\times 10^4$ epochs envisaged by our revised observation
strategy are mainly near quadrature, this would imply a photon-limited
trigonometric-parallax measurement of precision,
\begin{equation}
  \sigma(\pi_{\rel,c-s}) =
    4.6\,\mu{\rm as}\sqrt{10^{0.4(H_{{\rm Vega},c}-21)}+10^{0.4(H_{{\rm Vega},s}-21)}}.
  \label{eqn:par-prec}
\end{equation}
This could provide considerable additional discriminatory power to weed out
false candidates, depending on the brightness of the source and the candidates,
beyond the weeding done by the proper-motion comparisons.

To give a somewhat extreme but realistic example, suppose the actual
FFP had $M=M_\oplus$, $D_l = 2\,\kpc$, and $a_\perp = 10^4\,\au$,
i.e., similar to separations of the Solar System objects that feed the
long-period comets.  The host would be separated from the source by
$\Delta\theta = 5^{\prime\prime}$ inside of which there would be
roughly 300 ``candidates''.  In this example, suppose that
$D_s=8\,\kpc$ and $H_{{\rm Vega},s}=21$, while $M_{\rm
  host}=0.5\,M_\odot$ so that $H_{\rm Vega,host}=18$.  Then, the
astrometric measurement would be a random realization of
$\pi_{\rel,c-s} = 375\pm 5\,\mu$as, for example, $\pi_{\rel,c-s} =
381\pm 5\,\mu$as.  This astrometric measurement would be vetted
against the two possible $\pi_\rel$ values coming from the light-curve
analysis $\pi_{\rel,\pm} = \pi_{\e,\pm}\theta_\e$, which might be, for example,
$\pi_{\rel,+} = 410\pm 50\,\mu$as and $\pi_{\rel,-} = 270\pm 30\,\mu$as.
One would demand that any candidate be consistent with one of these two
at $2\,\sigma$, which would span
$210\,\mu{\rm as}<\pi_\rel< 510\,\mu{\rm as}$, corresponding approximately to
$3.0\,\kpc \ga D_l \ga 0.8\,\kpc$.  Of course, the actual host would easily
pass this vetting, being within $\sim 1\,\sigma$ of $\pi_{\rel,+}$.  However,
the great majority of other candidates would be removed by this cut
because only a small fraction of field stars seen toward the bulge lie within
$\leq 3\,\kpc$ of the Sun.  This is before vetting the vector proper motion
against the four possible values allowed by the light curve analysis.
In fact, the photon-limited astrometric precision was somewhat overkill in
this example, because the resulting error bar was 40 times smaller than
the range allowed by the light-curve prediction.  Hence, even a photometric
relative parallax would have been quite satisfactory.  Nevertheless, this
added precision would not ``go to waste'' (assuming it could be achieved)
because it would greatly improve the precision of the mass measurement.
The same would be true of the astrometric measurement of $\bmu_\rel$, which
(whether based on {\it Roman} or ELT astrometry)
would likely have much higher precision than the light-curve
based determination\footnote{Such higher precision is already routinely
achieved when the lens and source are separately resolved in late-time
high-resolution images, as in the specific cases of
OGLE-2005-BLG-071 \citep{ob05071c},
OGLE-2005-BLG-169 \citep{ob05169bat,ob05169ben},
MOA-2007-BLG-400 \citep{mb07400b},
MOA-2009-BLG-319 \citep{mb09319b},
OGLE-2012-BLG-0950 \citep{ob120950b}, and
MOA-2013-BLG-220 \citep{mb13220b}.
}.  Then the major contributor to the fractional mass error
would be (twice) the fractional error in $t_\e$ (from the light curve), via
$M= \mu_\rel^2 t_\e^2/\kappa\pi_\rel$.

In brief, a combination of light-curve data from {\it Roman} and a
second (i.e., parallax) observatory, late-time ELT astrometry and
photometry, and within-mission {\it Roman} astrometry can vet against
false-candidate hosts over a very wide range of separations, which
enable essentially unambiguous identification of all Unbound
FSPL FFPs, as well as excellent mass, distance, projected-separation,
and transverse-velocity measurements of all three categories of bound
FFPs.  The Unbound FFPs would then have mass and distance measurements
that were subject to the two-fold \citet{refsdal66} ambiguity, which
would have to be handled statistically.

\subsubsection{{Role of a Parallax Satellite for PSPL FFPs}
  \label{sec:parsat-pspl}}

A large fraction of low-mass FFPs will be FSPL simply because their
Einstein radii are small, so if the source suffers significant
magnification, it has a high probability to be transited by the FFP,
i.e., an FSPL event.  However, at higher masses, $M\ga M_\oplus$, a
declining fraction of FFP events will be FSPL.  For example, scaling to
typical values $\pi_\rel=50\,\mu$as and $\theta_*=0.3\,\mas$,
the fraction of FSPL events will be
\begin{equation}
  \eta_{\rm FSPL} = {\rm min}(1,\rho);\qquad \rho = 0.27
  \biggl({M \over M_\oplus}\biggr)^{-1/2}
  \biggl({\pi_\rel \over 50\,\mu\rm as}\biggr)^{-1/2}
  \biggl({\theta_* \over 0.3\,\mu\rm as}\biggr),
  \label{eqn:fsplfrac}
\end{equation}
so that, e.g., for $M>10\,M_\oplus$ (and for these fiducial parameters)
more than $92\%$ will be PSPL.  Because PSPL events generally do not
yield $\theta_\e$ measurements, it may first appear that
microlens-parallax measurements for PSPL FFPs would provide only
ambiguous information.

The lack of a $\theta_\e$ measurement is, in fact, the main issue 
for Unbound PSPL FFPs.  However, for bound PSPL
FFPs, the actual issue is host identification. If the host can be identified,
then the host-source relative proper motion can be measured astrometrically
(as was the case for bound FSPL FFPs) using either ELT or {\it Roman} data,
which will give a measurement of $\theta_\e$.

Thus, we focus in this section on the issue of host identification of
bound PSPL FFPs, under the assumption that they have $\bpi_\e$ measurements
(with, of course, \citealt{refsdal66} four-fold ambiguities).

The cases of Wide and Kuiper PSPL FFPs are very similar to their FSPL
counterparts that were discussed in Section~\ref{sec:parsat-fspl}.
There will actually be very few false candidates due the small offset
$\Delta\theta$ for these cases.  For FSPL FFPs, we vetted these only by
comparing the vector proper motion measurement from late-time
astrometry with the four values coming out of the light-curve
analysis.  For PSPL FFPs, we can compare only the {\it directions} of
the vector proper motions, but not their {\it amplitudes}.  However,
because of the small number of candidates, this should be adequate in
the great majority of cases.

As in the case of FSPL FFPs, the problem of confusion gradually worsens
as one moves toward higher $\Delta\theta$, and therefore in this regime
the loss of vetting from the amplitude of the proper motion may undermine some
identifications.

Eventually, within the Oort regime, $\Delta\theta$ becomes
sufficiently large that {\it Roman} field-star astrometry can be
brought into play.  In the context of FSPL FFPs, this enabled
simultaneous vetting by three parameters, i.e., a scalar plus a
two-vector: ($\pi_\rel,\bmu_\rel$).  The effect of removing the
$\theta_\e$ information is to reduce the vetting parameters from three
to two.  From the standpoint of astrometry, these can be expressed as
the ``projected velocity'', $\tilde \bv$,
\begin{equation}
  {\tilde \bv} = \au{\bmu_\rel\over\pi_\rel} .
  \label{eqn:vtilde}
\end{equation}
In fact, the ``projected velocity'' was originally introduced in a microlensing
context \citep{gould92}
as ${\tilde \bv} = ({\tilde r}_\e/t_\e) (\bpi_\e/\pi_\e)$, but we
can see that the two definitions are equivalent,
\begin{equation}
  {\tilde \bv} \equiv {{\tilde r}_\e\over t_\e}{\bpi_\e\over \pi_\e}
  = {{\tilde r}_\e\over t_\e}{\bmu_\rel\over \mu_\rel}
  = {{\tilde r}_\e\bmu_\rel\over \theta_\e}
  = {\au\over\pi_\e}{\bmu_\rel\over \theta_\e}
  = \au{\bmu_\rel\over\pi_\rel}.
  \label{eqn:vtilde2}
\end{equation}

Vetting with two parameters is clearly weaker than with 3 parameters.
It is premature to decide what to do about this in the absence of
real data, in particular, without an assessment of the robustness
of the astrometric measurement of $\tilde \bv$.  It may be, for example,
that the entire point is moot because there are extremely few Oort FFPs.
Or, it could be that $\tilde \bv$-based vetting works extremely well,
and there are no real issues of concern.  Or, it could be that there
are sufficiently many FSPL Oort FFPs at the higher masses $M\ga M_\oplus$
where PSPL predominates, that it is unnecessary to include the PSPL FFPs.
Or, most likely, the situation will be more complicated in some way
that we are unable to anticipate.  For the present, we content ourselves
with describing the vetting tools without forecasting how well they will
function in practice.

\section{{Additional Benefits}
  \label{sec:add-ben}}

As mentioned in Section~\ref{sec:intro}, re-orienting the {\it Roman}
observational strategy toward low-mass FFPs (by increasing the
cadence) will come at some cost to the total number of detected bound
planets, in particular those at relatively high mass ratios.
  For example, if {\it Roman} can support a cycle of 9
observations every 15 minutes, i.e., 36 observations per hour, then
these could be reorganized as
[$4\times\Gamma_8  + 1\times \Gamma_4$] 
rather than
[$9\times \Gamma_4$], where $\Gamma_n$ means ``$\Gamma= n\,{\rm hr}^{-1}$''.

This change would reduce the number of 2L1S bound planets over the entire
mass-ratio range $-4.5 < \log q < -1.5$, but by variable amounts,
with a greater reduction at the high mass end than the low-mass end.
We estimate that this reduction will be $\sim 25\%$ at $\log q=-4.5$
and $\sim 40\%$ at $\log q=-1.5$

However, this high-mass regime is
already reasonably well understood from the homogeneous KMT sample,
and will be much better understood by the time of {\it Roman} launch.
For example, the 2016-2019 sample contains
15 planets within the range $-4.5 < \log q \leq -4.0$, 
four planets within the range $-5.0 < \log q \leq -4.5$, and just one planet 
with $\log q \leq -5.0$.  It is plausible that the data already in hand
from 2021-2024 contain comparable numbers, and assuming that the experiment
continues through 2028, there will be an additional comparable number.
Thus, plausibly, there will be 45 KMT planets in the range
$-4.5 < \log q \leq -4.0$.  By a similar estimate, there will be of order
240 KMT planets in the range $-4.0 < \log q \leq -1.5$.  Thus, about
95 planets per dex over the three decades $-4.5 \leq \log q \leq -1.5$.
This implies roughly 10\% errors for each decade of mass ratio.

\subsection{{Lowest-mass-ratio 2L1S Planets}
  \label{sec:low-q}}

The real benefit from {\it Roman} will be to survey the regions $-5.5
< \log q \leq -4.5$ (where it will likely have relative sensitivities
similar to KMT at 1 dex higher, $-4.5 < \log q \leq -3.5$) and
especially $-6.0 < \log q \leq -5.5$ (where it will likely have
relative sensitivities similar to KMT at $-5.0 < \log q \leq -4.5$).
This is because, KMT has only weak sensitivity in the first of these
regions and essentially zero sensitivity in the second.  In the first
of these regions there will be little or no loss because the (4/9)
lost area will be partially or wholly compensated for by additional
planets recovered from higher cadence.  And in the second region,
there will be a net increase in detected planets, or at least in
planets that can be reliably characterized.

According to our understanding, {\it Roman} planet-parameter recovery has never
been simulated.  However, from our experience, this is a major issue
at low $q$, i.e., near the threshold of detectability.  That is, it
does little good to detect a planet that is actually $q=10^{-6}$, if
in the recovery, it is found to be equally likely to be $q=10^{-6}$
or $q=10^{-5.3}$.  This statement would not apply to a planet that was
found to be equally likely to be $q=10^{-4}$ or $q=10^{-3.3}$.  In that
case, one could assign Bayesian priors to each possibility based on the
hundred or so other planets whose mass ratio was reliably measured in
this regime.  But without a significant number of reliable $q=10^{-6}$
recoveries, there is no basis to establish reliable priors for ambiguous
cases.

Hence, for the lowest-$q$ planets, for which {\it Roman} will provide truly
unique information, doubling the cadence will enable more reliable recovery.

\subsection{{Low-mass  Wide-orbit 2L1S Planets}
  \label{sec:wide-2L1S}}

Physically, there is no distinction between wide planets in 2L1S events
and the Wide FFPs.  There is only the observational difference that
for the first, the host leaves traces on the microlensing event, whereas
for the second, it does not.  If the experiment had similar sensitivities
to both, then there would be far more Wide FFPs than wide 2L1S events
simply because (for planets at wide separation) a favorable geometry
is required for the host to leave a trace.

However, the sensitivity to wide planets in 2L1S events is potentially
much greater because the $\Delta\chi^2$ threshold is lower, perhaps
$\Delta\chi^2>60$ as opposed to $\Delta\chi^2>300$.  As discussed in
Section~\ref{sec:kuiper-special}, this is because the number of
effective trials is much less, roughly $10^9$ versus $10^{13}$.
Secondarily, for 2L1S events, $f_s$ is already known from the main event,
while for FFPs it must be determined either from the anomaly or from auxiliary
information.

The key point here is that the functional form of a wide 2L1S ``bump''
will differ very little from a Wide FFP ``bump'' at the same normalized
separation $s$, but which happens to lack a trace of the host due to
the geometry of the event.  This immediately begs the question of why,
if the FFP bumps require six 3-$\sigma$ points for proper characterization,
fewer would be needed to characterize the wide 2L1S bumps?  In fact, because
$f_s$ is already known (thus removing one degree of freedom), one could argue
that the requirement should be reduced from six to five.  Regardless of the
exact number, the origin of this ``requirement'' is not some arbitrarily
chosen selection criterion, but what is needed to have an interpretable event.
For wide low-mass 2L1S planets, with $\rho\ga 1$
(see Equation~(\ref{eqn:fsplfrac})),
the number of significantly magnified points is given by
Equation~(\ref{eqn:N-high-rho}).  Hence, for $\Gamma=4\,{\rm hr}^{-1}$ and
typical parameters, $N_{\rm exp}=3$, not 5 or 6.  By doubling the cadence
to $\Gamma=8\,{\rm hr}^{-1}$, these low-mass planets will be ``saved'', i.e.,
rendered interpretable.  Indeed, if one adopts the requirement of six points,
then the entire analysis of Section~\ref{sec:killing}, including
Figure~\ref{fig:comb}, can be directly applied.  If the requirement is
reduced to five points, then the analysis would be modified accordingly.

Again, we emphasize that the detection of these wide 2L1S planets can
be pushed down a factor $\sim 5$ in $\Delta\chi^2$ relative to the
physically similar but morphologically distinct Wide FFPs.  Thus,
while among physically wide planets, there will be many more Wide FFPs
than wide 2L1S planets, the latter have special importance because they
can probe to lower mass.

\subsection{{Auxiliary Science}
  \label{sec:auxiliary}}

The {\it Roman} microlensing survey will have many auxiliary science
applications.  Some of these have been studied in the literature.
Undoubtedly others will be identified only when the data are in hand.
These would all be impacted by a decision to double the cadence at the
expense of observing half the sky area.  These potential impacts
should be studied for each application separately.  Here we examine
a few applications in order to briefly argue that the science return
will generally be improved by making this change.

\subsubsection{{Transiting Planets}
  \label{sec:transits}}

{\it Roman} will be a powerful tool to detect transiting planets 
\citep{montet17,tamburo23,wilson23}.

In a transit study of this type, the $\Delta\chi^2$ threshold of detection is
set by the FAP  based on the effective number of trials, rather than
the signal required to characterize the planet.  One must consider
period steps, $\Delta P$, of $\Delta P= P/(\Gamma T)$ where $T=5\,$yr
is the duration
of the experiment.  For each $P$ and each diameter crossing time $\Delta t$,
one should consider $5\,P/\Delta t$ eclipse phases.  And for each of these,
perhaps 100 combinations of transit depth and impact parameter.  For
main-sequence stars with $R = R_\odot(M/M_\odot)$ radii, we have
$\Delta t/P = R/\pi a =
(R/R_\odot)(R_\odot/\au)/\pi [(M/M_\odot)(P/{\rm yr})^2]^{1/3}$, i.e.,
\begin{equation}
  {P\over\Delta t} = \pi{\au\over R_\odot}
  \biggl({P\over {\rm yr}}\biggr)^{2/3}
  \biggl({M\over M_\odot}\biggr)^{-2/3} = 
  13\biggl({P\over {\rm day}}\biggr)^{2/3}
  \biggl({M\over M_\odot}\biggr)^{-2/3}.
  \label{eqn:povert}
\end{equation}
And thus, for each observed star, one should consider a total of number of
trials,
\begin{equation}
  N_{\rm trial,star} = 500 \int_{P_\min}^{P_\max} {d P\over \Delta P}
  \,{P\over \Delta t} = 6500\int_{P_\min}^{P_\max} dP\,{\Gamma T\over P}
  \biggl({P\over {\rm day}}\biggr)^{2/3}
  \biggl({M\over M_\odot}\biggr)^{-2/3}\simeq
  10^4 \Gamma T \biggl({P_\max\over {\rm day}}\biggr)^{2/3}
  \biggl({M\over M_\odot}\biggr)^{-2/3}
  \label{eqn:ntrialstar}
\end{equation}
or
\begin{equation}
  N_{\rm trial,star} =
  10^{10} {\Gamma \rm\over 4\, hr^{-1}}\,{T\over 5\,\rm yr}
  \biggl({P_\max\over 10\,{\rm day}}\biggr)^{2/3}
  \biggl({0.6\,M\over M_\odot}\biggr)^{-2/3}.
  \label{eqn:ntrialstar2}
\end{equation}
Then, considering that there are of order $N_{\rm stars}\sim 10^8$ stars being
monitored, this yields a total number of trials
$N_{\rm trial,tot}=N_{\rm stars}N_{\rm trial,star} \sim 10^{18}$,
and thus a $\Delta\chi^2$ threshold
(assuming Gaussian statistics) of $\Delta\chi^2=2\ln N_{\rm trial,tot}=83$.
Plausibly, this should be increased by some amount to take account
of non-Gaussian noise, but this amount can only be determined from
having data in hand.

From the present standpoint, the key point is that the $\Delta\chi^2$
threshold depends on
$N_{\rm trial,tot} \propto \Gamma N_{\rm stars}$, so doubling $\Gamma$,
which automatically cuts $N_{\rm stars}$ by a factor 2,
does not impact the threshold.

Therefore doubling $\Gamma$ has the effect or reducing the planet-radius
detection threshold by a factor $2^{-1/4}\simeq 0.84$.  That is, the
noise remains essentially the same whether the planet is
transiting or not because only a small fraction of the light is occulted.
On the other hand, the signal is directly proportional to the planet area, $A$.
Therefore, $\Delta\chi^2 \propto \Gamma A^2$, i.e., $A_\min\propto \Gamma^{-1/2}$.

Thus, the effect of increasing $\Gamma$ will be to probe to 0.84 times
smaller planets at the expense of probing half as many potential hosts.
Even if we consider only hot Jupiters, for which the transit rate is of
order 0.1\%, there will be at least 100,000 transiting planets.  Thus, it
seems far more important to probe to the smallest planets possible, rather
than maximizing the total number detected.

\subsubsection{{Asteroseismology}
  \label{sec:seismic}}

\citet{gould15} argued that the {\it Roman} data stream could be mined
for asteroseismic signals in sources down to about $H_{\rm Vega} < 14$, of which
they estimated about $10^6$ would be in the {\it Roman} field.  Such
measurements can potentially yield the mass and radius of the sources,
although depending on the quality of the data, the two determinations
can be correlated.  However, assuming that photon-limited astrometry
can be extracted from the {\it Roman} data stream, the source radii can
be directly determined from a combination of their trigonometric parallaxes,
observed flux, observed color, and a color/surface-brightness relation.
The astrometric radii can then be cross checked with the asteroseismic
radii for the stars that are bright enough to have unambiguous asteroseismic
radii.  Assuming that the reliability of both are verified, the
astrometric radii can be used to constrain the asteroseismic solutions
of the fainter sources.

The effect of doubling $\Gamma$ will be, as usual, to improve the
measurements of each star at the expense of halving the number of stars.
To understand the first of these effects more quantitatively, we note
that from Figure~1 of \citet{gould15}, the flux error per observation scales as
$\sigma(\ln F)\propto F^{-1/3}$.
We can also state ``per observation'' as ``per 15 minutes'', according
to which doubling $\Gamma$ would decrease the ``per 15 minutes'' error
by $\sqrt{2}$.  That is, for a star at a given magnitude $H$, one would
achieve the same fractional error with the higher cadence, as one would
under the current regimen
for a star that is brighter by a factor $2^{3/2}$, i.e., $\Delta H = 1.1\,$mag.

\citet{gould15} estimate that the threshold of sensitivity begins just below
the clump, a region that is critical for probing stellar physics in the bulge,
a unique domain of {\it Roman}, compared to e.g., {\it Plato}
(2026 current launch date), which will target stars that are much closer to
Sun.  Given that the threshold lies near this key region of the color-magnitude
diagram, and considering the huge number of stars in the {\it Roman} sample,
it seems far more valuable to increase the S/N of each star, even at the
expense of losing half the area.

\subsubsection{{Kuiper Belt Objects}
  \label{sec:kuiperbelt}}

\citet{gould14b} argued that {\it Roman} could discover and measure
the orbits of about 5000 Kuiper Belt Objects (KBOs).  In contrast to
the other two applications that we have reviewed, the impact of
doubling $\Gamma$ on KBOs is somewhat complex.  In particular,
\citet{gould14b} estimated that 60\% of all KBOs that initially lay in
the {\it Roman} field (assumed to consist of 10 contiguous pointings)
would remain in it for the full 72-day season.  His main orbit
reconstruction calculations were restricted to this subsample.  He then
examined (his Figure~3) the effects of KBOs leaving and entering the
field and concluded that these effects are modest.

Naively, halving the number of fields would greatly decrease the fraction
of KBOs remaining in the field, possibly requiring much more detailed
calculations of the effect.  However, we believe that his original
estimate of only 60\% of the KBOs remaining in the field during the whole
72 days is probably in error.

There are two effects.  First, at $a=40\,\au$, the KBOs move in their
orbit at $v_{\rm orb} = (a/\au)^{-1/2} v_\oplus$ and therefore have a
proper motion $\mu=v_{\rm orb}/a = (1.4^\circ/{\rm yr})(a/40\,\au)^{-3/2}$ or
$\sim 0.28^\circ$ during a 72-day season.  Second, because the season
is approximately centered on quadrature, Earth moves back and forth by
$[1-\cos(2\pi\times 36\,{\rm day/yr})]\au = 0.19\,\au$, which yields a
reflex motion of $0.27^\circ (a/40\,\au)^{-1}$.  For (northern) spring
seasons, these two effects add at the beginning but are contrary at
the end, while the reverse is true of the autumn seasons.
Approximating the KBO and Earth orbits as circular, and focusing on
the spring season for definiteness, the instantaneous apparent motion
is $(v_\oplus/a)[(a/\au)^{-1/2}- \sin(2\pi t/{\rm yr})]$, which
reverses sign at $t = ({\rm
  yr}/2\pi)\sin^{-1}[(a/\au)^{-1/2}]\rightarrow 9.2\,$d after equinox
for $a=40\,\au$, at which point the relative displacement of Earth and
the KBO is $0.296\,\au$, or an angle of $0.42^\circ$.  Hence, if the
field were $2.8\,{\rm deg}^2$ and more-or-less square, the fraction
leaving (and possibly re-entering) the field would be about 25\%.
Thus, with the smaller field that we are proposing, the fraction
leaving would be about 40\%.  As \citet{gould14b} has already shown
that such a fraction does not have much impact, we conclude that the
increase in $\Gamma$ does not adversely affect orbit reconstruction.

However, it does still reduce the total number of KBOs in the field by
close to a factor of two, while increasing the effective depth of the
survey by doubling the number of measurements.  Because the measurements are
below sky, it is likely that the added S/N from the extra measurements would
be highly welcomed when analyzing the data.

Finally, we note that the change in cadence will have absolutely no
effect on the detection and measurement of KBO occultations.  This is because
the occultation time is short compared to the exposure time, so the
number of occultations is just proportional to the total number of pointings,
without reference to the specific pointing direction.

\subsection{{Possible Compromises}
  \label{sec:oldorder}}

As we argued in Section~\ref{sec:intro}, the principal revolutionary
potential of {\it Roman} microlensing lies in FFPs.   To fully exploit
this potential requires a radical revision of the observing strategy,
\begin{equation}
  [9\times \Gamma_4] \rightarrow [4\times\Gamma_8  + 1\times \Gamma_4], 
  \label{eqn:revision}
\end{equation}
where, again, $\Gamma_n$ means ``$\Gamma= n\,{\rm hr}^{-1}$''.

Moreover, the remaining revolutionary potential from the original
{\it Roman} 2L1S-centric microlensing program lies in extremely
low-mass planets, and these are also best pursued using the observing
strategy of Equation~(\ref{eqn:revision}).

Nevertheless (as is often the case), for reasons ranging from
bureaucratic intransigence, to ``treaty commitments'', to the
recalcitrance of outdated thinking, such a thoroughgoing FFP-centric
revolution may not be possible, at least not immediately.

Therefore, we describe several possible compromises.

\subsubsection{{$[9\times\Gamma_4]\rightarrow[2\times\Gamma_8+5\times\Gamma_4]$}
  \label{sec:comp1}}

This approach would target 7 fields, each with at least the $\Gamma_4$
cadence of the original strategy, so that none of the goals of the
original strategy would be qualitatively undermined.  At the same
time, it would permit testing of the new strategy.  Assuming that FFP
searches were carried out quickly, the results might argue for a
complete change of strategy, as described in
Equation~(\ref{eqn:revision}), after a year or two.  Or, failing that,
it could lay the basis for full adoption of
Equation~(\ref{eqn:revision}) in an extended mission.  In particular,
the existence of a large $\Gamma_8$ data stream would permit a direct
assessment of what would have been lost by reverting to a $\Gamma_4$
strategy, simply by masking every other data point.

\subsubsection{{$[9\times\Gamma_4]\rightarrow[1\times\Gamma_8+7\times\Gamma_4]$}
  \label{sec:comp2}}

This approach would be a more severe version of the compromise laid
out in Section~\ref{sec:comp1}.  It would drastically undermine the
practical results that would flow from Equation~(\ref{eqn:revision}), but
would enable testing of the concepts and advantages that we have outlined.

\subsubsection{{Summary Statement on Compromises}
  \label{sec:compall}}

Again, we do not advocate any of these (or other) compromises.  We believe
the case for an FFP-centric strategy is clear.  However, we also recognize
that in the real world, compromise must always be considered as an option.

\acknowledgments
We thank Zexuan Wu for assistance with ZAMS isochrone models.
J.C.Y. acknowledges support from US NSF Grant No. AST-2108414.
S.D. acknowledges the New Cornerstone Science Foundation through the XPLORER PRIZE.

\appendix
\section{{Primer on FFP Events and Mass Measurements}
\label{sec:primer}}

Here we give a comprehensive introduction to microlensing and microlensing
mass measurements, specifically with respect to FFPs.  As such, we will
restrict attention to single-lens single-source (1L1S) microlensing events.

A microlensing event consists of a lens of mass $M$ at distance $D_l$ and
source star of flux $f_s$ and radius $R_s$ at distance $D_s$.  Microlensing
events are primarily expressed in terms of lens-source relative astrometric
variables, which are then scaled to the Einstein radius, $\theta_\e$,
\begin{equation}
  \theta_\e \equiv  \sqrt{\kappa M\pi_\rel}=
  1.11\,\mu{\rm as}\biggl({M \over M_\oplus }\biggr)^{1/2}
  \biggl({\pi_\rel \over 50\,\mu\rm as }\biggr)^{1/2};
  \qquad \kappa = {4 G\over c^2\,\au} = 8144\,{\mu{\rm as}\over M_\odot}
  \label{eqn:thetae}
\end{equation}
The astrometric variables are the standard 5-parameter position, parallax,
proper motion (pppm) $(\btheta,\pi,\bmu)$, where $\btheta$ and $\bmu$
are two-vectors.  The relative lens-source astrometric variables are then
\begin{equation}
  (\btheta_\rel,\pi_\rel,\bmu_\rel) = (\btheta,\pi,\bmu)_l -
  (\btheta,\pi,\bmu)_s
  \label{eqn:relast}
\end{equation}
and the astrometric variables scaled to $\theta_\e$ are (or rather,
in a perfect world, would be)
\begin{equation}
  {\bf u}\equiv {\btheta_\rel\over\theta_\e};\qquad
  \pi_\e = {\pi_\rel\over \theta_\e};\qquad
  \bomega = {\bmu_\rel\over\theta_\e}.
  \label{eqn:scaled}
\end{equation}
However, in practice, the latter two variables are modified/replaced by
\begin{equation}
  t_\e = {1\over \omega};\qquad
  \bpi_\e = \pi_\e {\bmu_\rel\over \mu_\rel}.
  \label{eqn:tebpie}
\end{equation}
That is, whereas in astrometry, the proper motion is a vector, which
indicates the direction of motion, in microlensing, the proper motion
is a scalar (and is expressed inversely as a timescale, $t_\e$),
while the direction of motion is associated with the microlens
parallax $\bpi_\e$.  Finally, the source radius is also expressed as
an angle, $\theta_*$ which is also scaled to the $\theta_\e$,
\begin{equation}
  \rho\equiv{\theta_*\over\theta_\e};\qquad
  \theta_* = {R_s\over D_s}.
  \label{eqn:rho}
\end{equation}

If the parallactic reflex motion of Earth can be ignored (as
is almost always the case for FFPs), then the normalized
trajectory is given by
\begin{equation}
  {\bf u}(t) = {\bf u}_0 + \bomega(t-t_0) =
  {\bf u}_0 + {(t-t_0)\over t_\e}{\bmu_\rel\over\mu_\rel}; \qquad
  ({\bf u}_0 \cdot \bmu_\rel \equiv 0),
  \label{eqn:bfu}
\end{equation}
implying (by the Pythagorean theorem),
\begin{equation}
  u(t) = \sqrt{{(t-t_0)^2\over t_\e^2} + u_0^2},
  \label{eqn:pyth}
\end{equation}
where $t_0$ is the time of closest approach and ${\bf u}_0$ is the
closest position (vector impact parameter).  If $u_0\gg\rho $, i.e.,
the lens passes well outside the face of the source, then the flux evolution
is unaffected by the finite-size of the source
(point-source point-lens [PSPL] event) and is given by
\begin{equation}
  f(t) = f_s A(t) + f_b; \qquad
  A(t;t_0,u_0,t_\e) = {u^2 + 2\over u\sqrt{u^2 + 4}},
  \label{eqn:aot}
\end{equation}
where $A$ is the magnification and
$f_b$ is the blended light that does not participate in the event.
One can then solve for the five parameters $(t_0,u_0,t_\e,f_s,f_b)$ from
the light curve.  The Einstein timescale, 
\begin{equation}
  t_\e = 
  {\theta_\e\over \mu_\rel}= 
  {\sqrt{\kappa M\pi_\rel}\over \mu_\rel} =
  1.6\,{\rm hr}\,\biggl({M \over M_\oplus}\biggr)^{1/2}
  \biggl({\pi_\rel \over 50\,\mu\rm as}\biggr)^{1/2}
  \biggl({\mu_\rel \over 6\,\masyr}\biggr)^{-1},
  \label{eqn:te2}
\end{equation}
then gives an indication of $M$ (``short events have low mass'') but
does not determine it because the actual mass further scales inversely
with the unknown $\pi_\rel$ and quadratically with the unknown $\mu_\rel$.

This problem can be partially solved if the lens transits the face of the
source $(\rho\ga u_0)$ in which case the magnification is a function
of four variables $A(t;t_0,u_0,t_\e,\rho)$, i.e., a finite-source point-lens
(FSPL) event.  Then, $\theta_\e=\theta_*/\rho$ can be determined provided
that $\theta_*$ is known.  There are standard techniques \citep{ob03262}
for measuring $\theta_\e$ from microlensing data.  As discussed in
Section~\ref{sec:challenges}, these can break down for {\it Roman} FFPs,
but as discussed in Section~\ref{sec:solution}, $\theta_\e$ can usually
be recovered even when they do.

When $\theta_\e$ is measured, $M$ is better constrained because it only
depends on one unknown variable $(\pi_\rel)$, but it is still not
unambiguously determined.  For this, it is necessary to measure either
$\pi_\rel$ itself, or the microlens parallax, $\bpi_\e$.  In the latter
case, both $\pi_\rel$ and the lens mass $M$ are determined:
\begin{equation}
  \pi_\rel = \theta_\e\pi_\e;\qquad M = {\theta_\e\over\kappa\pi_\rel}.
  \label{eqn:mpirel}
\end{equation}
The techniques for measuring $\pi_\rel$ and/or $\bpi_\e$ are extensively
discussed in Sections~\ref{sec:parallax} and \ref{sec:integrated}, and
(with one exception) we do not repeat that discussion here.

The exception is the use of adaptive optics (AO) on extremely large
telescopes (ELTs) to measure the mass of bound planets.  The only way
to determine whether an FFP (by definition, an event for which there
is no light-curve evidence of a host) is bound, is to search for its
putative host in late-time AO observations.  If a host is found, then
$\bmu_\rel$ can easily be measured from two epochs, and so
$\theta_\e = \mu_\rel t_\e$ of the planet can also be determined, even
if the event is PSPL.  Then, there can be several possible routes
to measuring the lens mass.  First, by combining this measurement of
$\bmu_\rel$ with the flux of the host, one can measure the distance
(also the mass) of the host \citep{ob05169bat,ob05169ben,masada},
and thereby obtain an estimate of $\pi_\rel$.  Then,
$M_{\rm planet}=\theta_\e^2/\kappa\pi_\rel$.  Second, $\pi_\rel$ of the host
and source can be measured directly from astrometry.  Third, if there
are L2-parallax measurements of $\bpi_\e$, then these can be combined
with $\theta_\e$ (and also the directional information from $\bmu_\rel$)
to yield a unique mass.

Some typical values of quantities mentioned here are $D_s \sim 8\,\kpc$,
$D_l\sim 4$--$6\,\kpc$, $\pi_\rel = 50\,\mu$as for disk lenses and
$\pi_\rel = 15\,\mu$as for bulge lenses, $\mu_\rel \sim 6\,\masyr$,
$\theta_* \sim 0.3$--$0.6\,\mu$as.  Equations~(\ref{eqn:thetae}) and
(\ref{eqn:te2}) give the values of $\theta_\e$ and $t_\e$ when scaled
to these typical parameters.  Note that $\rho\ga 1$
(i.e., $\theta_\e\la \theta_*$) for $M\la 0.5\,M_\oplus$.

\input tabzeta

\clearpage

\begin{figure}
\plotone{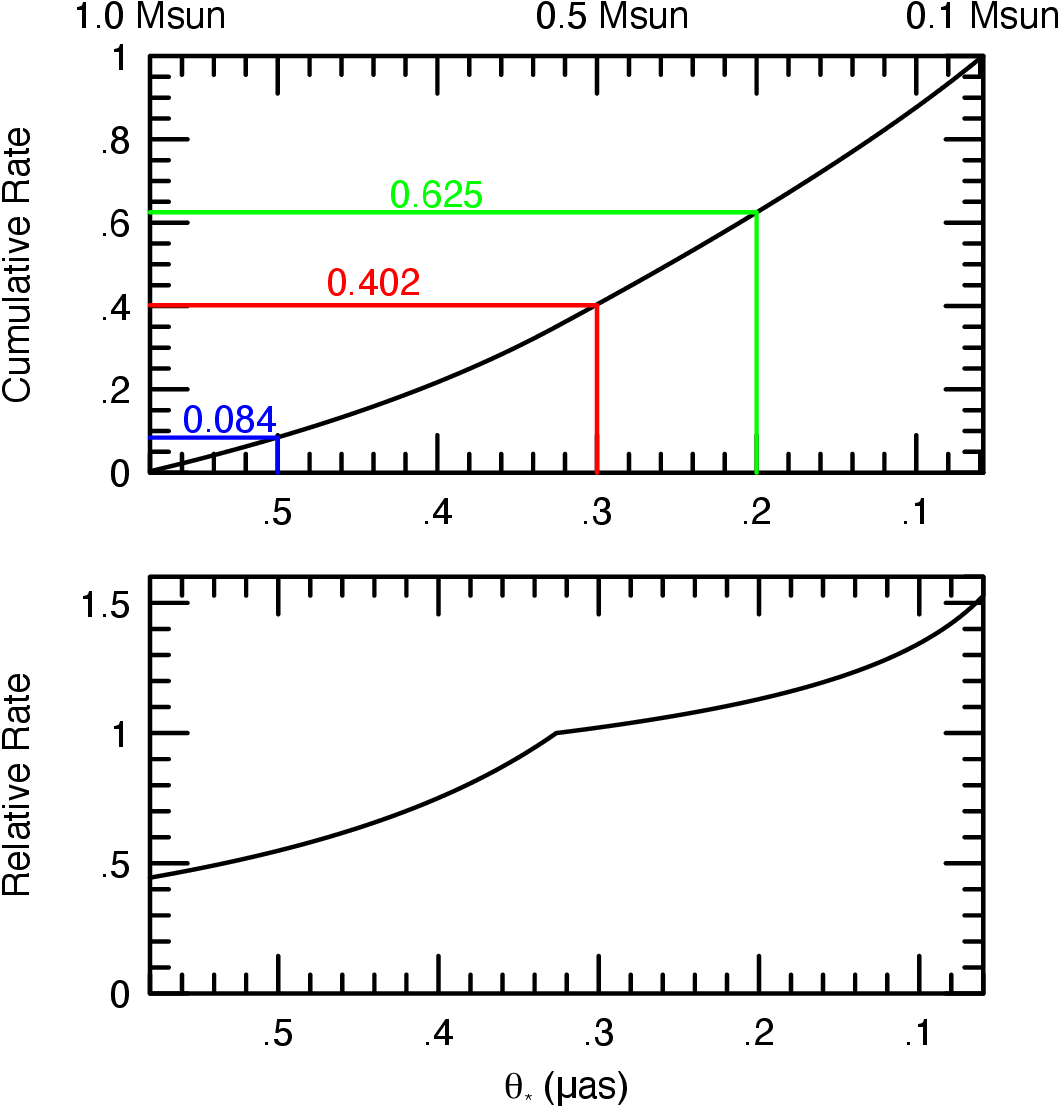}
\caption{Differential and cumulative $\theta_*$ distributions
  based on the \citet{sweeps} mass function, taking account that for
  $\rho\gg 1$, the event cross section is $\propto\theta_*$, and that
  on the main sequence, $\theta_*\propto M_s$.  Bigger $\theta_*$ is
  better because the number of magnified observations scales $N_{\rm
    exp}\propto \theta_*$.  See Equation~(\ref{eqn:N-high-rho}), which
  scales to $\theta_*=0.3\,\mu$as, so that $\theta_*\geq 0.6\,\mu$as
  would be required for an FFP detection if all other factors were
  unchanged.  In practice, at the current $\Gamma=4\,{\rm hr}^{-1}$, detections
  come from both lower $\mu_\rel$ (Figure~\ref{fig:cum-mu}) and higher
  $\theta_*$.  See Section~\ref{sec:role-gamma} and Figure~\ref{fig:comb}.
}
\label{fig:thetastar}
\end{figure}

\begin{figure}
\plotone{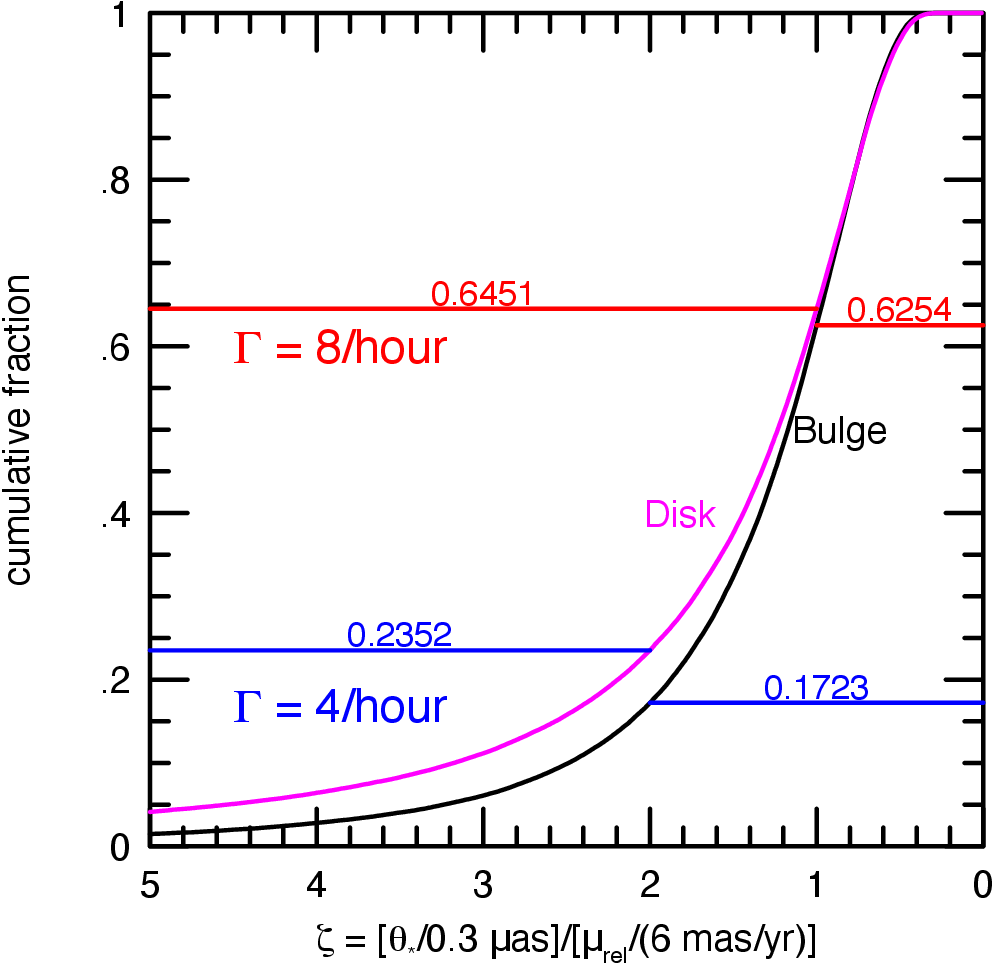}
\caption{Cumulative distribution of large-$\rho$ FFP events that satisfy
  the $N_{\rm exp}=6$ requirement, taking account of both the distributions
  of $\mu_\rel$ (Figure~\ref{fig:cum-mu}) and $\theta_*$
  (Figure~\ref{fig:thetastar}).  The abscissa is the fraction of events
  with ratios $\theta_*/\mu_\rel$ that exceed a certain given factor ($\zeta$)
  relative to the scaling factors in Equation~(\ref{eqn:N-high-rho}),
  whose prefactor is 3.0.  Thus, 
  to meet the $N_{\rm exp}=6$ requirement with $\Gamma = 4\,{\rm hr}^{-1}$,
  the factor $\zeta$ must be at least 2.0, but with $\Gamma = 8\,{\rm hr}^{-1}$,
  a factor of $\zeta\geq 1.0$ is sufficient.  The ratio of surviving events 
  is $(0.6254/0.1723)= 3.63$ for bulge lenses and $(0.6451/0.2352)= 2.74$
  for disk lenses.  Weighting these by the FSPL bulge/disk ratio
  of 2.5:1 (Figure~9 from \citealt{gould22}) gives an overall average of 3.38.
}
\label{fig:comb}
\end{figure}

\begin{figure}
\plotone{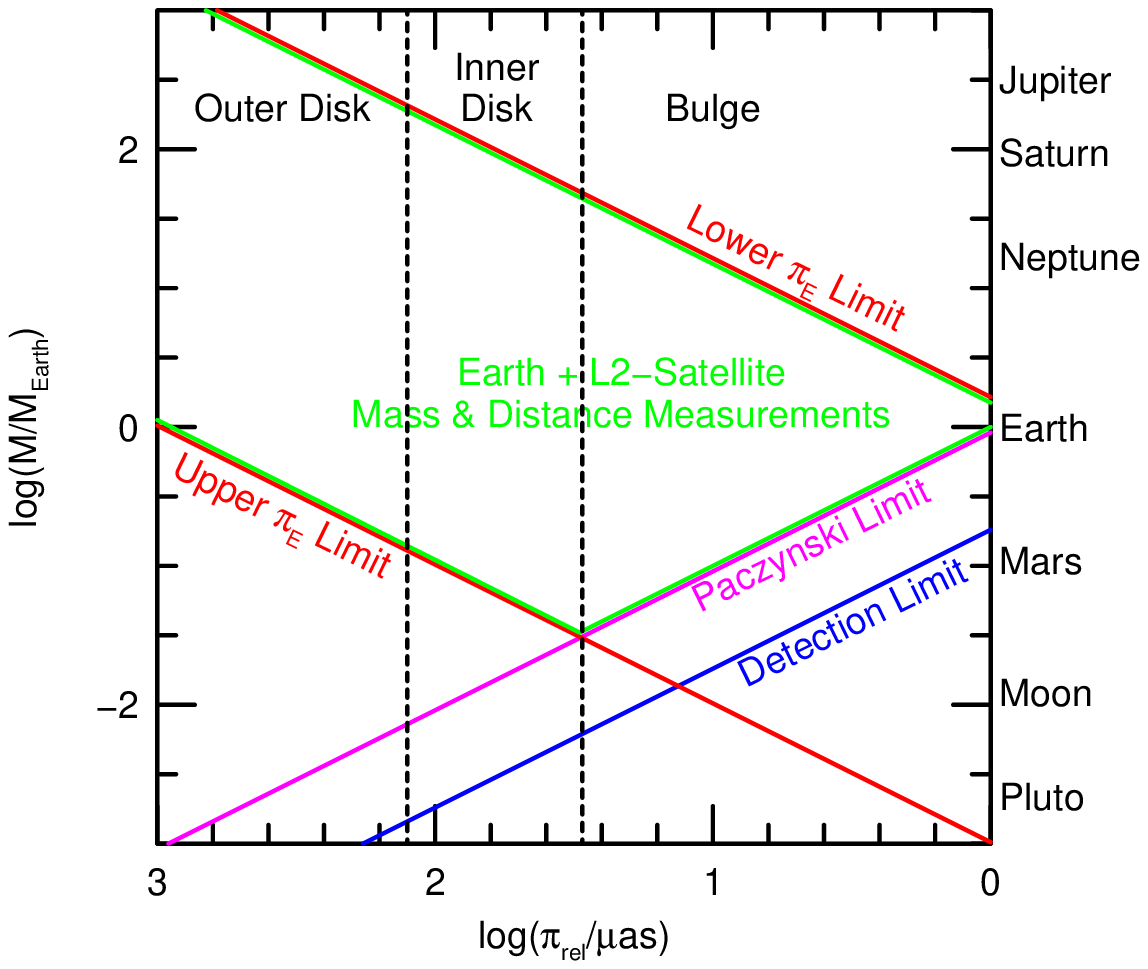}
\caption{Region of the $(\log \pi_\rel,\log M)$ plane that is
  accessible to microlens-parallax measurements (green), as
  constrained by the Refsdal Limits (red) on the microlens parallax
  ($\pi_\e=\sqrt{\pi_\rel/\kappa M}$) and the Paczy\'nski Limit
  (magenta) on the Einstein radius ($\theta_\e=\sqrt{\kappa
    M\pi_\rel}$).  The first requires that the Earth-L2 projected
  separation, $D_\perp = 0.01\,\au$, lies in the range $0.05\la
  D_\perp/{\tilde r}_\e \la 2$ relative to the projected Einstein
  radius ${\tilde r}_\e\equiv \au/\pi_\e$, i.e., big enough that the
  Earth and the satellite see sufficiently different events to make a
  measurement, yet small enough that both observatories actually see a
  signal.  The second requires that $\theta_\e$ is large enough that
  the \citet{pac86} parameters $(u_0,t_\e)$ can be measured from the
  light curve.  Some FFPs (with $\rho=\theta_*/\theta_\e>2$) fall below
  the latter threshold, but can still be detected (blue).  These can
  have non-parallax mass measurements provided that they are bound and
  their hosts can be identified.  The masses of various Solar-System
  bodies are shown for reference.  Regions corresponding to the
  outer-disk ($D_l\la 4\,\kpc$), inner-disk ($4\,\kpc \la D_l\la
  6.5\,\kpc$), and bulge ($D_l\ga 6.5\,\kpc$) FFPs are delineated by
  dashed lines.  }
\label{fig:par}
\end{figure}

\end{document}

%% file: author.tex
\author{\textsc{
Andrew Gould$^{1,2}$
Jennifer C.\ Yee$^{3}$,
Subo Dong$^{4,5}$
}}

\affil{$^{1}$Department of Astronomy, Ohio State University, 140 W.
18th Ave., Columbus, OH 43210, USA}

\affil{$^{2}$Max-Planck-Institute for Astronomy, K\"{o}nigstuhl 17,
69117 Heidelberg, Germany}

\affil{$^{3}$ Center for Astrophysics $|$ Harvard \& Smithsonian, 60 Garden
St., Cambridge, MA 02138, USA}

\affil{$^{4}$Department of Astronomy, School of Physics, Peking University, Yiheyuan Rd. 5, Haidian District, Beijing, China, 100871}

\affil{$^{5}$Kavli Institute for Astronomy and Astrophysics, Peking University, Yi He Yuan Road 5, Hai Dian District, Beijing 100871, China}

%% file: tabzeta.tex
 \begin{deluxetable}{rlllllll}
 \tablecolumns{8} \rotate \tablewidth{0pc}
 \tablecaption{\textsc {Complement to Figure~\ref{fig:comb}}}
 \tablehead{\colhead{$\Gamma\ ({\rm hr}^{-1})$} & 
\colhead{$\zeta$} &
\colhead{Cum (bulge)} &
\colhead{Cum (disk)} &
\colhead{ratio (bulge)} &
\colhead{ratio (disk)} &
\colhead{weighted ratio ($W$)} &
\colhead{$W\zeta/2$} }

  \startdata
     2 & 4.00 & 0.0282 & 0.0643 & 0.164 & 0.273 & 0.195 & 0.390 \\
{\bf 4} & {\bf 2.00} & {\bf 0.1723} & {\bf 0.2352} & {\bf 1.000} & {\bf 1.000} & {\bf 1.000} & {\bf 1.000} \\
     6 & 1.33 & 0.4044 & 0.4500 & 2.347 & 1.913 & 2.223 & 1.482 \\
{\bf 8} & {\bf 1.00} & {\bf 0.6254} & {\bf 0.6451} & {\bf 3.630} & {\bf 2.743} & {\bf 3.377} & {\bf 1.689} \\
  10 & 0.80 & 0.7853 & 0.7878 & 4.558 & 3.347 & 4.212 & 1.685 \\
  12 & 0.67 & 0.8847 & 0.8811 & 5.135 & 3.375 & 4.632 & 1.544 \\
\hline
 \enddata
 \tablecomments{Bold-faced lines are illustrated in Figure~\ref{fig:comb}.}
 \label{tab:zeta}
 \end{deluxetable}